\documentclass[10pt,notitlepage,aps,pre]{revtex4-1}

\usepackage{graphicx}
\usepackage{amsmath}
\usepackage{color}

 \usepackage[colorlinks,hyperindex]{hyperref}
 \usepackage{hyperref}
	\hypersetup
	{
		colorlinks,%
		citecolor=blue,%
		linkcolor=blue,%
		urlcolor=black,%
	}

\newcommand{\tr}{\mbox{tr}}

\begin{document}

\title{Strain tensor selection and the elastic theory of incompatible thin sheets}

\author{Oz Oshri} 
\email{ozzoshri@tau.ac.il} 
\affiliation{Raymond \& Beverly Sackler School of Physics \& Astronomy, Tel Aviv University, Tel Aviv 6997801, Israel}

\author{Haim Diamant} 
\email{hdiamant@tau.ac.il} 
\affiliation{Raymond \& Beverly Sackler School of Chemistry, Tel Aviv University, Tel Aviv 6997801, Israel}

\date{25 November 2016}

\begin{abstract}
The existing theory of incompatible elastic sheets uses the deviation
of the surface metric from a reference metric to define the strain
tensor [Efrati et al., J.\ Mech.\ Phys.\ Solids {\bf 57}, 762
  (2009)]. For a class of simple axisymmetric problems we examine an
alternative formulation, defining the strain based on deviations of
distances (rather than distances squared) from their rest
values. While the two formulations converge in the limit of small
slopes and in the limit of an incompressible sheet, for other cases
they are found not to be equivalent. The alternative formulation
offers several features which are absent in the existing theory. (a)
In the case of planar deformations of flat incompatible sheets, it
yields linear, exactly solvable, equations of equilibrium. (b) When
reduced to uniaxial (one-dimensional) deformations, it coincides with
the theory of extensible elastica; in particular, for a uniaxially
bent sheet it yields an unstrained cylindrical configuration. (c) It
gives a simple criterion determining whether an isometric immersion of
an incompatible sheet is at mechanical equilibrium with respect to
normal forces. For a reference metric of constant positive Gaussian
curvature, a spherical cap is found to satisfy this criterion except
in an arbitrarily narrow boundary layer.
\end{abstract}

\maketitle

\section{Introduction}

In the past two decades there has been a renewed interest in the
elasticity of thin solid sheets in view of the wealth of surface
patterns and three-dimensional (3D) shapes that they exhibit under
stress
\cite{cerda3,huan10,Davidovitch11,vand11,cerda4,brau13,king12,audo11,cerda5}.
In addition, experiments and models have been devised for {\it
  incompatible} sheets, which contain internal residual stresses even
in the absence of external forces
\cite{marder,sharon,Efi1,Efi2,Efi3,Efi4,klein2,Efi5,dervaux,marder2,pezzulla,dervaux2,santangelo,Dias,audoly3,guven,muller2,BenAmar20052284,Goriely2,Lewicka402}. The
study of such sheets has been motivated by their relevance to
morphologies in nature \cite{marder,marder2,dervaux,dervaux2,Armon}
and frustrated self-assembly \cite{Armon,doronPRL}. Incompatible
sheets form nontrivial 3D shapes {\em spontaneously}. They can also be
``programmed" to develop a desired 3D shape
\cite{klein2,klein,Dias,naomi,kim,bae}.

The necessary existence of sheets with unremovable internal stresses
is rationalized as follows. When treating a thin solid sheet as a
mathematical surface, its relaxed state is characterized by a 2D
reference metric tensor, $\bar{g}_{\alpha\beta}$, associated with the
relaxed in-plane configuration, and a reference second fundamental
form, $\bar{b}_{\alpha\beta}$, related to the relaxed out-of-plane
configuration (curvature) \cite{Efi1}. (We shall use Latin indices
$(i,j,\ldots)$ for 3D coordinates and Greek indices
$(\alpha,\beta,\ldots)$ for 2D ones.) However, not any
$\bar{g}_{\alpha\beta}$ and $\bar{b}_{\alpha\beta}$ correspond to a
physical surface. For the surface to be embeddable in 3D Euclidean
space, these forms must satisfy a set of geometrical constraints
\cite[p.~203]{lipschutz}. Thus, in general, an actual sheet will be
incompatible\,---\,its actual metric and second fundamental form,
$a_{\alpha\beta}$ and $b_{\alpha\beta}$, will not coincide with their
reference counterparts\,---\,leading to unavoidable intrinsic
stresses.

A covariant theory for incompatible elastic bodies has been presented
by Efrati, Sharon, and Kupferman (referred to hereafter as ESK)
\cite{Efi1} and successfully applied to several experimental systems
\cite{Armon,moshe2,doronPRL}. Their elastic energy for a 3D body
reads,
\begin{eqnarray}
&&E_{\rm 3D}=\int_{\mathcal{V}}\mathcal{A}^{ijkl}\tilde{\epsilon}_{ij}\tilde{\epsilon}_{kl}\sqrt{|\bar{g}|}dV,  \nonumber
 \\ 
&&\tilde{\epsilon}_{ij}=\frac{1}{2}(g_{ij}-\bar{g}_{ij}), \label{ESD-3D}
\end{eqnarray}
where the integration is over the unstrained volume, $\mathcal{V}$,
$g_{ij}$ and $\bar{g}_{ij}$ are the metric and reference metric,
$\bar{g}$ is the determinant of the reference metric and
$\mathcal{A}^{ijkl}$ is the elastic tensor. To explicitly distinguish
the strain used by ESK we mark it with a tilde. ESK also presented a
dimensional reduction of this energy to 2D for incompatible thin
elastic sheets, resulting in a sum of stretching and bending
contributions,
\begin{eqnarray}
 \text{ESK:} \ \ \ E_{\rm 2D}&=&E_s+E_b=\frac{t}{2}\int_{A} \mathcal{A}^{\alpha\beta\gamma\delta}\tilde{\epsilon}_{\alpha\beta}\tilde{\epsilon}_{\gamma\delta}\sqrt{|\bar{g}|}dA+\frac{t^3}{24}\int_{A} \mathcal{A}^{\alpha\beta\gamma\delta}b_{\alpha\beta}b_{\gamma\delta}\sqrt{|\bar{g}|}dA,
\label{ESK-2D} \nonumber \\
\ \ \ \tilde{\epsilon}_{\alpha\beta}&=&\frac{1}{2}(a_{\alpha\beta}-\bar{g}_{\alpha\beta}), \label{Green-strain}
\end{eqnarray}   
where $t$ is the sheet thickness, the integral is over the unstrained area, and
$\tilde{\epsilon}_{\alpha\beta}$ is the ESK two-dimensional strain tensor.

Arguably, the functional in Eq.~(\ref{ESD-3D}) represents the simplest
covariant theory of incompatible elasticity. It makes a certain choice
of strain tensor, which is based on the relative deviations of the
distances {\it squared} from their rest values (the so-called
Green--St.~Venant strain tensor \cite{Efi1,Fu,Dill}). In elasticity
theory the strain measure is regarded as a parametrization
freedom\,---\,so long as the stress tensor (and resulting energy
functional) is appropriately defined, different definitions of the
strain tensor will lead to the same equilibrium deformation of the
elastic body \cite[Sec.~2.5]{Dill}. Indeed, other choices of strain
have been made in compatible elasticity, such as the Biot strain
tensor \cite{Biot}, which expresses the spring-like deviations of
distances within the body.  Generally, one can write a dimensionless
deviation of a certain variable $\ell$ from its reference $\ell_0$ as
$\Delta=\frac{1}{m\ell_0^m}\left(\ell^m -\ell_0^m\right)$, where $m$
is an arbitrary number \cite[p.~6]{Fu}. In the limit of small
deviations, $\Delta\ll 1$, one always gets $\Delta\simeq
(\ell-\ell_0)/\ell_0$ for any $m$. Thus, it seems that within linear
elasticity of infinitesimal strains the choice of $m$ is immaterial.

Dimensional reduction of 3D linear elasticity to 2D thin sheets
introduces non-quadratic terms in the reduced energy functional. As we
shall see below, a different selection of the strain tensor for the 3D
body\,---\,the incompatible analogue of Biot's strain\,---\,leads to
non-quadratic terms in 2D which differ from those obtained from
Eq.\ (\ref{Green-strain}). Thus, the resulting theory is not
equivalent to the ESK one. This holds even in the case of a compatible
sheet with a flat reference metric  \cite{Koiter,ciarlet}. The differences between the two
formulations are quantitatively small but have a qualitative effect on
the structure of the theory and the simplicity of its application. 
We note that the present work is not the first to indicate the effect of 
strain-tensor selection. Similar observations were made in the
context of compatible beam theory \cite{Irschik}.


We begin in Sec.~\ref{2D-alt-model} by presenting the alternative
formulation based on Biot's selection of 3D strain. We perform a
reduction to 2D, which is limited to axisymmetric surface deformations
along the principal axes of stress. In Sec.~\ref{cylindrical-symmetry}
we apply the formulation to the simple example of a compatible sheet
that is uniaxially bent by boundary moments. We show that it
coincides in this case with the extensible elastica, yielding a bent,
unstrained, cylindrical shape, whereas the choice made in
Eq.~(\ref{ESK-2D}) gives a cylinder with non-zero in-plane strain.
Section~\ref{exact-sol-planar} presents further applications to
several examples of incompatible flat discs. We derive linear
equations of equilibrium, and obtain their analytical solutions, for
problems which are described by nonlinear equations in the ESK theory.
Section~\ref{self-consis} presents a self-consistency criterion, based
on the alternative formulation, for the stability of axisymmetric
isometric immersions of such discs with respect to internal bending
moments. We apply the criterion to the case of a reference metric with
constant positive Gaussian curvature, whose isometric immersion is a
spherical cap. In Sec.~\ref{conclusions} we conclude and discuss
future extensions of this work.

\section{Alternative two-dimensional formulation for simple deformations}
\label{2D-alt-model}

We impose three requirements on the alternative formulation for 2D
incompatible sheets: (a) It should be invariant under rigid
transformations (rotations and translations). (b) In the limit of
incompressibe compatible sheets it should converge to the known
Willmore functional \cite{Efi1}. (c) In the small-slope approximation
it should converge to the F\"oppl-von K\`arm\`an (FvK) theory
\cite{Landau}.

The formulation presented here holds for a small subset of problems
which we can treat exactly. We consider a disc-like thin sheet of
radius $R$, and parametrize it by the polar coordinates
$(r,\theta)$. The relaxed length, squared, of a line element on the
sheet is given by the following reference metric,
\begin{equation}
\bar{g}_{\alpha\beta}=
\begin{pmatrix}
1 & 0 \\ 0 & \Phi^2(r)
\end{pmatrix}
\ \ \ , \ \ \ ds^2=dr^2+\Phi^2(r)d\theta^2,
\label{ref-metric}
\end{equation}
where $dr$ is the relaxed arclength element along the radial direction
and $2\pi\Phi(r)$ is the relaxed perimeter of a circle of radius $r$
around the disc center. Once $\Phi(r)\neq r$ the flat configuration
contains internal strains. While such a sheet may have a complicated
equilibrium deformation, we restrict ourselves to surfaces of
revolution. The 3D position of a displaced point on the surface is
given by
\begin{equation}
{\bf f}(r,\theta)=[r+u_r(r)]{\bf \hat{r}}+\zeta(r){\bf \hat{z}},
\label{f-shape}
\end{equation}
where $u_r$ is the radial displacement, $\zeta$ is the height
function, $\hat{{\bf r}}$ is a unit vector tangent to the sheet in the
radial direction, and $\hat{{\bf z}}$ is a unit vector in the
perpendicular direction to the flat disc. Note that, for an
incompatible sheet, the case of $u_r(r)=\zeta(r)=0$ does not
correspond to a stress-free configuration.

The 2D energy functional of this system can be derived out of a 3D
formulation using the Kirchhoff-Love hypothesis
\cite{Love,Goldenveizer,Libai,Efi1,Koiter,Dias}. For this purpose we
identify the 2D sheet defined above with the mid-surface of a 3D slab.
Under the Kirchhoff-Love set of assumptions the configuration of the
3D body is given by,
\begin{equation}\label{dfstar}
{\bf f}^{\star}(r,\theta,x_3)={\bf f}(r,\theta)+x_3 \hat{{\bf n}}(r,\theta),
\end{equation}
where $x_3\in [-t/2,t/2]$ is a coordinate  in the direction ${\bf \hat{n}}$ normal to the mid-surface,
\begin{equation}
{\bf \hat{n}}=\frac{\partial_r{\bf f}\times\partial_{\theta}{\bf f}}{|\partial_r{\bf f}\times\partial_{\theta}{\bf f}|}=\frac{(1+\partial_r u_r){\bf \hat{z}}-\partial_r\zeta{\bf \hat{r}}}{\sqrt{(1+\partial_r u_r)^2+(\partial_r\zeta)^2}}.
\end{equation}
On a surface of constant $x_3$, the length squared of an infinitesimal
line element is found, after some algebra, to be,
\begin{equation}
d{\bf f^{\star}}^2=\left[a_{rr}-2x_3 b_{rr}+x_3^2c_{rr}\right]dr^2+\left[a_{\theta\theta}-2x_3 b_{\theta\theta}+x_3^2 c_{\theta\theta}\right] d\theta^2,
\label{df-star-square}
\end{equation} 
where $a_{\alpha\beta}=\partial_{\alpha}{\bf
  f}\cdot\partial_{\beta}{\bf f}$,
$b_{\alpha\beta}=-\partial_{\alpha}{\bf f}\cdot \partial_{\beta}
\hat{{\bf n}}$, and $c_{\alpha\beta}=\partial_{\alpha}\hat{{\bf
    n}}\cdot\partial_{\beta}\hat{{\bf n}}$, are the first, second, and
third fundamental forms.

On the other hand, following Biot's approach \cite[p.~17]{Biot}, a
pure deformation of that surface is represented by the symmetric
transformation matrix,
\begin{equation}
\begin{pmatrix}
dr' \\ \Phi d\theta'
\end{pmatrix}
=
\begin{pmatrix}
1+\epsilon^{\star}_{rr} & \epsilon^{\star}_{r\theta} \\ \epsilon^{\star}_{r\theta} & 1+\epsilon^{\star}_{\theta\theta}
\end{pmatrix}
\begin{pmatrix}
dr \\ \Phi d\theta
\end{pmatrix}
\label{def-Biot},
\end{equation}
where $\epsilon_{\alpha\beta}^{\star}$ is the in-plane strain tensor
of the constant-$x_3$ surface. Note that 
this definition of the strain
correponds to changes in length (not length squared). Thus,
\begin{eqnarray} 
d{\bf f^{\star}}^2&=&dr'^2+(\Phi d\theta')^2 \nonumber \\&=&  \left[(1+\epsilon^{\star}_{rr})^2+(\epsilon^{\star}_{r\theta})^2\right]dr^2+\left[(1+\epsilon^{\star}_{\theta\theta})^2+(\epsilon^{\star}_{r\theta})^2\right](\Phi d\theta)^2+2\epsilon^{\star}_{r\theta}(2+\epsilon^{\star}_{rr}+\epsilon^{\star}_{\theta\theta})\Phi d\theta dr.
\label{general-trans}
\end{eqnarray}
Comparinging Eqs.~(\ref{df-star-square}) and (\ref{general-trans}), we
identify,
\begin{subequations}\label{epsilon-star}
\begin{eqnarray}
&&\epsilon_{rr}^{\star}=\sqrt{(1+\epsilon_{rr})^2-2x_3b_{rr}+x_3^2c_{rr}}-1, \label{epsilon-rr-star} \\ 
&&\epsilon_{\theta\theta}^{\star}=\sqrt{(1+\epsilon_{\theta\theta})^2-2x_3 b_{\theta\theta}/\Phi^2+x_3^2 c_{\theta\theta}/\Phi^2}-1, \label{epsilon-2theta-star} \\
&&\epsilon_{r\theta}^{\star}=0, \label{epsilon-rtheta-star}
\end{eqnarray}
\end{subequations}
where
\begin{subequations}\label{epsilon-in-plane}
\begin{eqnarray}
&&\epsilon_{rr}=\sqrt{a_{rr}}-1=\sqrt{(1+\partial_r u_r)^2+(\partial_r \zeta)^2}-1, \label{epsilon-rr} \\
&&\epsilon_{\theta\theta}=\sqrt{a_{\theta\theta}}/\Phi-1=\frac{r}{\Phi}-1+\frac{u_r}{\Phi}. \label{epsilon-2theta} 
\end{eqnarray}
\end{subequations}
We have reached a definition of the mid-surface in-plane strains in
terms of the actual and reference metrics, based on the spring-like
deformed length rather than length squared.

\begin{figure}[!tbh]
\vspace{0.7cm}
  \centering
  \includegraphics[width=5cm]{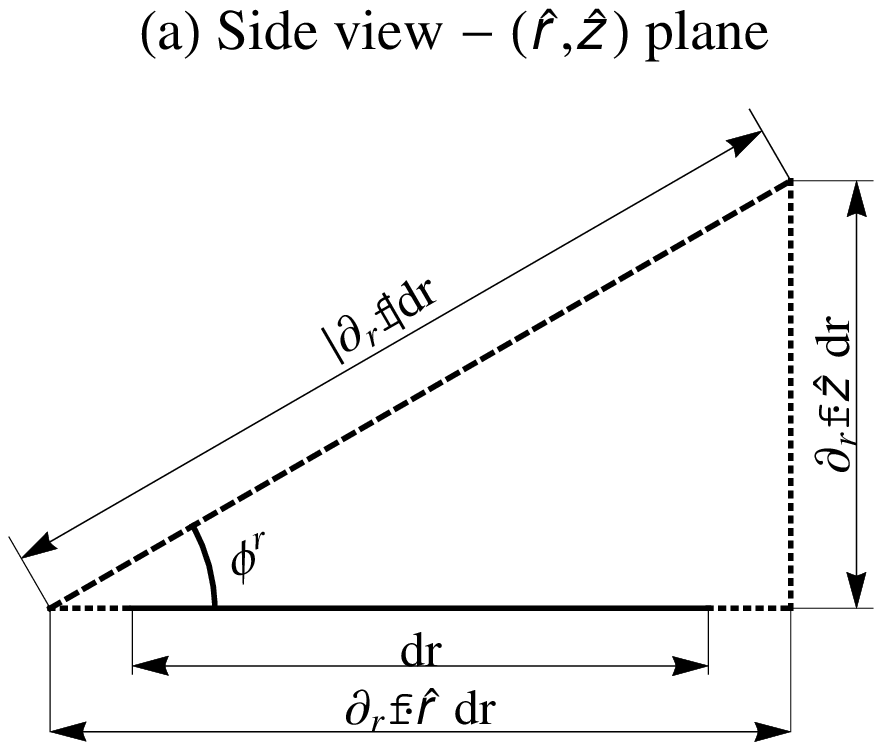} \  \  \ \ \
	\includegraphics[width=5cm]{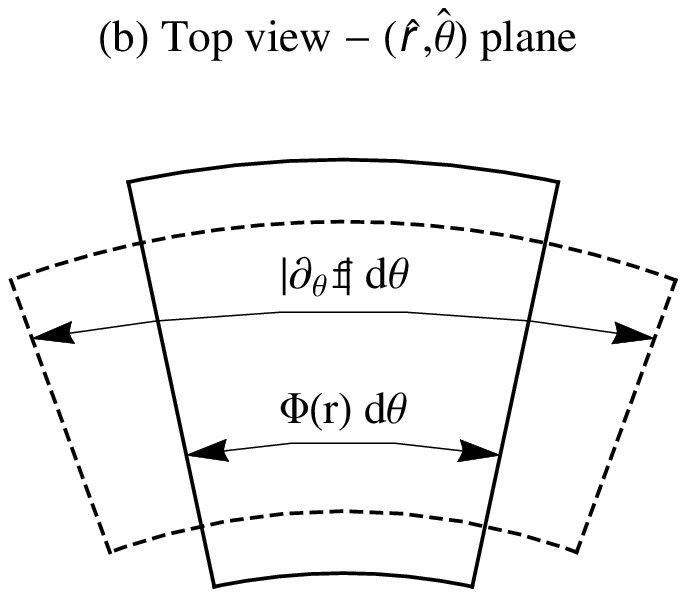}
	\includegraphics[width=4cm]{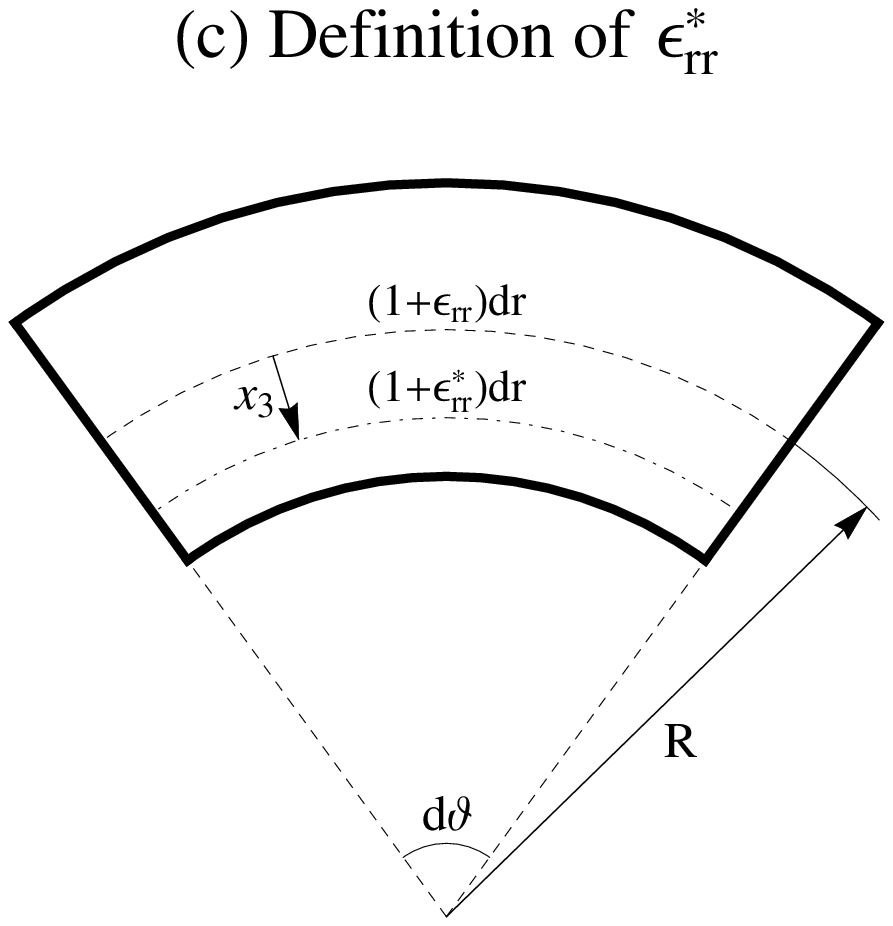}
	\caption{(a) Deformation of an infinitesimal element in the
          radial direction. The relaxed length of the element is $dr$
          (solid line), and the deformed length is $|\partial_r{\bf
            f}|dr$. The radial strain component is
          $\epsilon_{rr}=\frac{|\partial_r{\bf f}|dr-dr}{dr}$, as
          given by Eq.~(\ref{epsilon-rr}). The angle $\phi^r$
          satisfies $\sin\phi^r=\partial_r {\bf f}\cdot {\bf
            \hat{z}}/|\partial_r{\bf f}|$.  Substituting ${\bf
            f}(r,\theta)$ from Eq.~(\ref{f-shape}) in the latter
          relation and using Eq.~(\ref{phi-2theta}) gives
          $\phi_{\theta\theta}=\sin\phi^r/\Phi$. In addition, by
          direct differentiation it can be verified that
          $\phi_{rr}=\partial_r\phi^r$ as given by
          Eq.~(\ref{phi-rr}). (b) Deformation of an infinitesimal
          sheet element in the azimuthal direction. The relaxed length
          in this direction is $\Phi d\theta$ (solid line) and the
          deformed length is $|\partial_{\theta}{\bf f}|d\theta$
          (dashed line). Thus, the azimuthal strain is
          $\epsilon_{\theta\theta}=\frac{|\partial_{\theta}{\bf
              f}|d\theta-\Phi d\theta}{\Phi d\theta}$, as given by
          Eq.~(\ref{epsilon-2theta}). (c) Deformation of an
          infinitesimal line element in the radial direction at height
          $x_3$ below the mid-surface. By geometry, the shown angle 
          $d\vartheta=(1+\epsilon_{rr})dr/R=(1+\epsilon_{rr}^{\star})dr/(R-x_3)$. Using
          $1/R=(1+\epsilon_{rr})^{-1}d\phi^r/dr$ and solving for
          $\epsilon_{rr}^{\star}$ gives Eq.~(\ref{epsilon-rr-star2}).
        }
  \label{epsilon-rr-2theta}
\end{figure}

The geometrical interpretation of these strains is illustrated in
Fig.~\ref{epsilon-rr-2theta}. The fact that the strains describe
deformed lengths \cite[p.~41]{Goldenveizer} leads at this stage to two
simplifications. First, the fundamental forms satisfy the simple
relations, $c_{rr}=b_{rr}^2/(1+\epsilon_{rr})^2$ and $\Phi^2
c_{\theta\theta}=b_{\theta\theta}^2/(1+\epsilon_{\theta\theta})^2$. Second,
once these expressions are substituted in Eqs.~(\ref{epsilon-star}),
we can rewrite the strains at constant $x_3$ as,
\begin{subequations}\label{epsilon-star-simplified}
\begin{eqnarray}
&&\epsilon_{rr}^{\star}=\epsilon_{rr}-x_3\phi_{rr}, \label{epsilon-rr-star2} \\ 
&&\epsilon_{\theta\theta}^{\star}=\epsilon_{\theta\theta}-x_3\phi_{\theta\theta}. \label{epsilon-2theta-star2} 
\end{eqnarray}
\end{subequations}
(See Fig.~\ref{epsilon-rr-2theta}c for the geometrical meaning of these strains.) Here we have defined the out-of-plane strains,
\begin{subequations}\label{bending-strains}
\begin{eqnarray}
&&\phi_{rr}=\sqrt{c_{rr}}=\frac{(1+\partial_r u_r)\partial_{rr}\zeta-\partial_{rr} u_r \partial_r\zeta}{(1+\partial_r u_r)^2 +(\partial_r\zeta)^2}, \label{phi-rr}\\
&&\phi_{\theta\theta}=\sqrt{c_{\theta\theta}}/\Phi=\frac{1}{\Phi}\frac{\partial_r\zeta}{\sqrt{(1+\partial_r u_r)^2+(\partial_r\zeta)^2}}. \label{phi-2theta}
\end{eqnarray}
\end{subequations}
Defining further $\phi^r$ and $\phi^{\theta}$ as the tangent angles in the radial and azimuthal directions of the surface of revolution, we find $\phi_{rr}=\partial_r\phi^r$ and $\phi_{\theta\theta}=(1/\Phi)\partial_{\theta}\phi^{\theta}$ (see Fig.~\ref{epsilon-rr-2theta}(a) and the explanation in its caption). 
This clarifies the geometrical meaning of the  ``bending-strains", $\phi_{rr}$ and $\phi_{\theta\theta}$.

In the framework of linear elasticity the energy functional of the 3D
slab is given by \cite{Landau},
\begin{equation}
E_{\rm 3D}=\frac{E}{2(1-\nu^2)}\int_{-t/2}^{t/2}\int_0^{R}\int_0^{2\pi}\left[(\epsilon_{rr}^{\star})^2+(\epsilon_{\theta\theta}^{\star})^2+2\nu\epsilon_{rr}^{\star}\epsilon_{\theta\theta}^{\star}\right]\Phi d\theta dr dx_3,
\label{E-z-dependent}
\end{equation} 
where $E$ is Young's modulus and $\nu$ the Poisson ratio. Substituting
Eqs.~(\ref{epsilon-star-simplified}) in (\ref{E-z-dependent}) and
integrating over $x_3$ gives,
\begin{equation}
E_{\rm 2D} = 
  \frac{Y}{2}\int_0^{R}\int_0^{2\pi}\left[\epsilon_{rr}^2+\epsilon_{\theta\theta}^2
  +2\nu\epsilon_{rr}\epsilon_{\theta\theta}\right]\Phi d\theta
dr+\frac{B}{2}\int_0^{R}\int_0^{2\pi}\left[\phi_{rr}^2+\phi_{\theta\theta}^2+2\nu\phi_{rr}\phi_{\theta\theta}\right]\Phi
d\theta dr,
\label{E-2D-final}
\end{equation}
where $Y=Et/(1-\nu^2)$ is the stretching modulus and $B=Et^3/12(1-\nu^2)$ is the bending modulus. The first integral in Eq.~(\ref{E-2D-final}) is the stretching energy,
\begin{equation}
E_s=\frac{1}{2}\int_0^R\int_0^{2\pi}\left[\sigma_{rr}\epsilon_{rr}+\sigma_{\theta\theta}\epsilon_{\theta\theta}\right]\Phi d\theta dr,
\label{Es-final-sigma}
\end{equation}
where  the stress components  $\sigma_{\alpha\beta}=\delta E/\delta \epsilon_{\alpha\beta}$ are given by,
\begin{subequations}\label{sigma-2D}
\begin{eqnarray}
&&\sigma_{rr}=Y(\epsilon_{rr}+\nu\epsilon_{\theta\theta}), \label{sigma-rr1} \\
&&\sigma_{\theta\theta}=Y(\epsilon_{\theta\theta}+\nu\epsilon_{rr}). \label{sigma-2theta1} 
\end{eqnarray}
\end{subequations}
Similarly, the second integral in Eq.~(\ref{E-2D-final}) gives the bending energy,
\begin{equation}
E_b=\frac{1}{2}\int_0^R\int_0^{2\pi}\left[M_{rr}\phi_{rr}+M_{\theta\theta}\phi_{\theta\theta}\right]\Phi d\theta dr,
\label{Eb-final-M}
\end{equation}
where the bending moments,  $M_{\alpha\beta}=\delta E/\delta \phi_{\alpha\beta}$, in the radial and azimuthal directions are given by,
\begin{subequations}\label{M-2D}
\begin{eqnarray}
&&M_{rr}=B(\phi_{rr}+\nu\phi_{\theta\theta}), \label{M-rr1} \\
&&M_{\theta\theta}=B(\phi_{\theta\theta}+\nu\phi_{rr}) \label{M-2theta1}.
\end{eqnarray}
\end{subequations}

Looking back at the dimensional reduction performed, we see why a generalization from axisymmetric deformations to general ones, although possible, is going to be much more cumbersome. 

Let us now verify that the three requirements that we have imposed on
the energy functional are fulfilled by Eq.~(\ref{E-2D-final}). The
first requirement, of invariance under rigid transformations, is
satisfied, since the strains have been derived from a pure deformation
matrix, Eq.~(\ref{def-Biot}), as discussed in the first chapter of
Ref.~\cite{Biot}. Equivalently, Eqs.~(\ref{Es-final-sigma}) and
(\ref{Eb-final-M}) can be rewritten in terms of the tensor invariants,
\begin{eqnarray}
&&E_s= \frac{Y}{2}\int_0^R\int_0^{2\pi}\left[\tr({\epsilon})^2-2(1-\nu)\det{(\epsilon)}\right]\sqrt{|\bar{g}|}d\theta dr,\nonumber \\
&&E_b= \frac{B}{2}\int_0^R\int_0^{2\pi}\left[\tr(\bar{g}^{-1}c)+2\nu\sqrt{\det{(\bar{g}^{-1}c)}}\right]\sqrt{|\bar{g}|}d\theta dr, \nonumber
\end{eqnarray}
which is manifestly invariant to rigid transformations. 
To verify the second requirement, we take the incompressible limit,
$a_{\alpha\beta}\rightarrow \bar{g}_{\alpha\beta}$, and obtain
$E_s=0$, $\phi^2_{rr}\rightarrow \kappa_{rr}^2$ and
$\phi^2_{\theta\theta}\rightarrow \kappa_{\theta\theta}^2$, where
$\kappa_{rr}$ and $\kappa_{\theta\theta}$ are the two principal
curvatures on the surface in the radial and azimuthal
directions. Substituting the latter relations in the second integral
of Eq.~(\ref{E-2D-final}), we obtain,
\begin{equation}
\text{Incompressible sheet:} \ \ \ E_b=\frac{B}{2}\int_0^R\int_0^{2\pi}\left((\kappa_{rr}+\kappa_{\theta\theta})^2-2(1-\nu)\kappa_{rr}\kappa_{\theta\theta}\right)\Phi dr d\theta,
\label{Eb-Willmore}
\end{equation}
which coincides with the known Willmore functional
\cite{Efi1}. Lastly, we verify the third requirement, that for
compatible sheets in the small-slope approximation our model converges
to the FvK theory \cite{Landau}. Setting $\Phi=r$ and expanding the
in-plane strain, Eqs.~(\ref{epsilon-in-plane}), to linear order in
$u_r$ and quadratic order in $\zeta$, we have $\epsilon_{rr}\simeq
\partial_r u_r +\frac{1}{2}(\partial_r \zeta)^2$ and
$\epsilon_{\theta\theta}=u_r/r$. The latter strains along with
Eq.~(\ref{Es-final-sigma}) yield the stretching energy in the FvK
approximation \cite{benny1}.  Similarly, the ``bending strains",
Eqs.~(\ref{bending-strains}), are approximated by $\phi_{rr}\simeq
\partial_{rr}\zeta$ and $\phi_{\theta\theta}\simeq
\partial_r\zeta/r$. Substituting these in Eq.~(\ref{Eb-final-M}), we
obtain the FvK bending energy,
\begin{equation}
\text{Small slope:} \ \ \ E_b\simeq\frac{B}{2}\int_0^R\int_0^{2\pi}\left[(\nabla_r^2\zeta)^2-2(1-\nu)[\zeta,\zeta]\right]r dr d\theta,
\label{Eb-energy-f-FvK1}
\end{equation}
where $\nabla_r^2\zeta\equiv\frac{1}{r}\partial_r(r\partial_r\zeta)$ and $[\zeta,\zeta]\equiv\frac{1}{r}\partial_r\zeta\partial_{rr}\zeta$ are the small-slope approximations of the mean and Gaussian curvatures. 

\section{Uniaxial deformation by bending}
\label{cylindrical-symmetry}

\begin{figure}[!tbh]
\vspace{0.1cm}
\centering
\includegraphics[width=10cm]{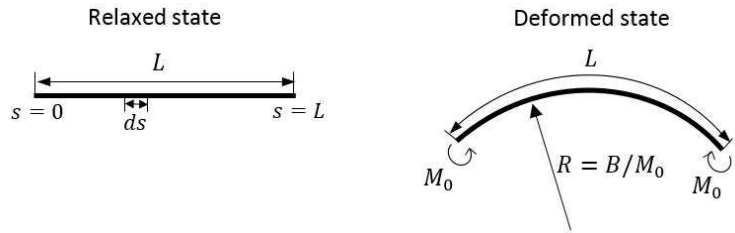}
\caption{A flat thin sheet is deformed into a cylinder of constant
  radius without stretching of its midplane. This deformation is
  obtained for the extensible elastica by applying bending moments, $M_0$, on
  the sheet edges or by imposing $d\phi/ds$ at the boundaries.}
\label{cylindrical-deformation}
\end{figure}

We would like to demonstrate the difference between the ESK model and
the one presented in the preceding section, using the simplest example
possible. Consider the uniaxial deformation of a compatible sheet by
bending moments applied at its edges. Alternatively, we can replace
the moments by purely geometrical boundary conditions on the
configuration at the edges, as given below. Since no in-plane axial
forces are applied, a particularly simple possibility is a purely bent
cylindrical deformation of the sheet's midplane\,---\,an isometry
which contains no stretching energy
(Fig.~\ref{cylindrical-deformation}). Indeed, this is the deformation
obtained in this case from the theory of extensible elastica
\cite{Reissner1972,magnusson,analogy,humber}, as we recall below.

To apply the formulation to this simple problem we should reduce the
2D energy, Eq.~(\ref{E-2D-final}), to 1D. Consider a radial cut of a
$\theta$-independent deformation as a planar compatible filament
($\Phi(r)=1$). Identify $r\rightarrow s$, where $s\in[0,L]$ is the
undeformed arclength along the filament, and $\phi^r(r)\rightarrow
\phi(s)$, the angle between the tangent to the filament and the flat
reference plane. We then have
$\phi_{rr}^2\rightarrow\phi_{ss}^2=(d\phi/ds)^2$,
$\epsilon_{rr}\rightarrow \epsilon_{ss}$, and
$\phi_{\theta\theta}=\epsilon_{\theta\theta}=0$. Substitution of these
relations in Eq.~(\ref{E-2D-final}) gives,
\begin{equation}
  E_{\rm 1D}=E_s+E_b=\int_0^L\left[\frac{Y}{2}\epsilon_{ss}^2+\frac{B}{2}\left(\frac{d\phi}{ds}\right)^2\right]ds.
\label{E-1D-no-P}
\end{equation}
This functional coincides with the energy of an extensible elastic
filament in a planar deformation as given by the theory of extensible
elastica \cite{Reissner1972,magnusson,analogy,humber}.  

Alternatively, we could reduce the sheet into a filament through an
azimuthal cut along a narrow annulus of large radius $\rho$, in which
case $d\theta\rightarrow ds$,
$\phi_{\theta\theta}^2\rightarrow\phi_{ss}= (d\phi/(\Phi ds))^2$, and
$\phi_{rr}=\epsilon_{rr}=0$.  We then obtain
\begin{equation}
E_{\rm
  1D}=E_s+E_b=\int_0^{L'}\left[\frac{Y}{2}\epsilon_{ss}^2+\frac{B}{2}\left(\frac{d\phi}{\Phi
    ds}\right)^2\right]\Phi ds.
\label{E-1D-gss}
\end{equation}
The parameter $s$ now runs between $0$ and $L'$, such that
$L=\int_0^{L'}\Phi ds$ is the total relaxed length. In addition,
$\epsilon_{ss}$ now measures the in-plane strain with respect to the
prescribed metric. 
The energy of Eq.~(\ref{E-1D-gss}) is the extension of the extensible
elastica theory to the case of a nontrivial reference metric.

Returning to the ordinary elastica, we note that Eq.~(\ref{E-1D-no-P})
can be derived from a discrete model of springs and joints
\cite{analogy} while enforcing from the outset the decoupling between
the stretching and bending contributions \cite[p.~77]{Libai}. In
Eq.~(\ref{E-1D-no-P}) this decoupling is manifest in the independence
of $E_s$ on $\phi$, $\frac{\delta E_s}{\delta\phi}=0$, while $E_b$ is
independent of $\epsilon_{ss}$, $\frac{\delta E_b}{\delta
  \epsilon_{ss}}=0$.  In the absence of boundary axial forces, the
equations of equilibrium are obtained from minimization of
Eq.~(\ref{E-1D-no-P}). Defining the in-plane stress (acting to only
locally stretch the filament) and bending moment (acting only to
change its local angle) as,
\begin{subequations}
\label{sigmaM}
\begin{eqnarray}
&&\sigma_{ss}\equiv\frac{\delta E_{\rm 1D}}{\delta \epsilon_{ss}}=Y\epsilon_{ss}, \label{sigma-ss-1D-def} \\
&&M_{ss}\equiv\frac{\delta E_{\rm 1D}}{\delta \left(\frac{d\phi}{ds}\right)}=B \frac{d\phi}{ds}, \label{M-ss-1D}
\end{eqnarray}
\end{subequations}
those equations of equilibrium are,
\begin{subequations}\label{1D-eqn-M-sigma}
\begin{eqnarray}
\sigma_{ss}=0,&& \label{sigma-ss-1D} \\
\frac{dM_{ss}}{ds}=0. \label{dM-ss-1D}&&
\end{eqnarray}
\end{subequations}

When a constant moment, $M_0$, is applied at the boundaries
(Fig.~\ref{cylindrical-deformation}),
Eqs.~(\ref{sigma-ss-1D-def})--(\ref{dM-ss-1D}) yield $\epsilon_{ss}=0$
and $\phi(s)=\phi(0)+(M_0/B) s$. This solution corresponds to a
circular arc of radius $B/M_0$ and total length $L$. Alternatively, if
we impose $(d\phi/ds)|_{s=0}=c$, we get $\phi(s)=\phi(0)+c s$, corresponding
to a circular arc of radius $1/c$. The energy of this configuration is
$E_{\rm 1D}=(B/2)c^2 L$.

The strain-free cylindrical shape is preserved also in the more
complicated case of a nonuniform reference metric,
Eq.~(\ref{E-1D-gss}).  Variation of this energy with respect to
$\epsilon_{ss}$ and $\phi$ gives, as before,
Eqs.~(\ref{1D-eqn-M-sigma}), where the in-plane stress is given again
by Eq.~(\ref{sigma-ss-1D-def}). The bending moment is modified to,
\begin{equation}
M_{ss}=\frac{\delta E_{\rm 1D}}{\delta \left(\frac{1}{\Phi}\frac{d\phi}{ds}\right)}=\frac{B}{\Phi}\frac{d\phi}{ds},
\label{M-ss-Phi}
\end{equation}
which replaces Eq.~(\ref{M-ss-1D}). The in-plane strain (with respect
to the reference metric) vanishes. When we apply a moment $M_0$ at the
boundaries, or impose $(d\phi/(\Phi ds))|_{s=0}=c$, we find again a
strain-free cylindrical shape with radius $B/M_0$, or $1/c$.

We now show that the ESK functional gives a different result.  We
specialize Eq.~(\ref{ESK-2D}) to the case of a compatible sheet under
uniaxial deformation. Since the deformation has zero Gaussian
curvature, we set $\bar{g}_{ss}= 1$ and, from
Eq.~(\ref{Green-strain}), obtain $a_{ss}=1+2\tilde{\epsilon}_{ss}$. In
addition, we have $\sqrt{|\bar{g}|}dA\rightarrow ds$,
$t\mathcal{A}^{ssss} \rightarrow Y$, and $\frac{t^3}{12}
\mathcal{A}^{ssss} \rightarrow\ B$. Substituting these relations in
Eq.~(\ref{ESK-2D}) gives
\begin{equation}
\text{ESK:} \ \ \ E_{\rm 1D}=\int_0^L\left(\frac{Y}{2}\tilde{\epsilon}_{ss}^2
  +\frac{B}{2} b_{ss}^2\right)ds.
\label{ESK-reduced}
\end{equation}
The relations between the variables appearing in the ESK
Eq.~(\ref{ESK-reduced}) and the ones in Eq.~(\ref{E-1D-no-P}) are
$\tilde\epsilon_{ss}=\epsilon_{ss}(1+\epsilon_{ss}/2)$, and
$b_{ss}=\partial_s(\sqrt{a_{ss}}\hat{{\bf t}})\cdot \hat{{\bf
    n}}=(1+2\tilde{\epsilon}_{ss})^{1/2}\frac{d\phi}{ds}$.  

Naively, if we set the variations of the energy (\ref{ESK-reduced})
with respect to $\tilde\epsilon_{ss}$ and $b_{ss}$ to zero, we will
get the same result as above, i.e., a strain-free circular
configuration with $\tilde\epsilon_{ss}=0$,
$b_{ss}=(d\phi/ds)_{s=0}=c$, and energy $E_{\rm 1D}=(B/2)c^2 L$. Thus,
the coupling between $\tilde\epsilon_{ss}$ and $d\phi/ds$ appearing in
$b_{ss}=(1+2\tilde{\epsilon}_{ss})^{1/2}\frac{d\phi}{ds}$ would not
have an effect on the configuration. However, the correct minimization
is with respect to the filament's trajectory ${\bf f}(s)$. As shown in
Appendix~\ref{app-1D-minimization}, this is equivalent to the
minimization with respect to $\epsilon_{ss}$ and $\phi$. In terms of
these variables, Eq.~(\ref{ESK-reduced}) becomes
\begin{equation}
\text{ESK:} \ \ \ E_{\rm
  1D}=\int_0^L\left[\frac{Y}{2}[\epsilon_{ss}(1+\epsilon_{ss}/2)]^2
+\frac{B}{2} [1+2\epsilon_{ss}(1+\epsilon_{ss}/2) \left(
  \frac{d\phi}{ds}\right)^2 \right]ds.
\label{ESK-reduced-2}
\end{equation}
The bending contribution to this energy depends on $\epsilon_{ss}$,
which results in a strained configuration under the boundary
conditions given above. Specifically, minimization of the energy in
Eq.~(\ref{ESK-reduced-2}) with respect to $\epsilon_{ss}$ and $\phi$,
under the boundary condition $(d\phi/ds)_{s=0}=c$, yields a circular
arc, $\phi(s)=\phi(0)+cs$, which nonetheless contains non-zero strain,
$\epsilon_{ss}= \sqrt{1-2Bc^2/Y}-1$. The energy of this configuration
is $E_{\rm 1D}=(B/2)c^2L[1-(B/Y)c^2]$, slightly deviating from the
energy of the extensible elastica obtained above.

Two comments should be added concerning the difference between the two
models. (a) As demonstrated by the case of a geometrical boundary
condition on $d\phi/ds$, the difference does not arise from different
definitions of the boundary bending moment. (This remains correct if
we impose the condition on the {\it apparent} curvature,
$[d\phi/d(1+\epsilon_{ss})s]_{s=0}$.) (b) In Ref.~\cite{Efi1} a term
proportional to $\tilde{\epsilon}_{ss}(d\phi/ds)^2$ was neglected in
the final step. Clearly, its inclusion merely changes the numerical
coefficient in the second term of Eq.~(\ref{ESK-reduced-2}).

In summary, unlike the formulation of Sec.~\ref{2D-alt-model}, the ESK
model does not strictly reduce to the extensible elastica. Under
uniaxial bending at the boundaries it produces a small in-plane
strain, while our formulation and the extensible elastica predict a
strain-free cylindrical shape. The discrepancy is small and vanishes
in the incompressible limit of $B/Y\rightarrow 0$. Moreover, the
correction terms are of order $(B/Y)c^2\sim (tc)^2$, which must always
be small in any elasticity theory of sheets of finite
thickness. Nevertheless, the effect of the coupling between stress and
bending moments goes beyond this simple 1D example and profoundly
affects the structure of the theory, as will be shown in the following
sections.

\section{Exact solutions for planar deformations of incompatible sheets}
\label{exact-sol-planar}

\begin{figure}[!tbh]
\vspace{0.7cm}
  \centering
  \includegraphics[width=7cm]{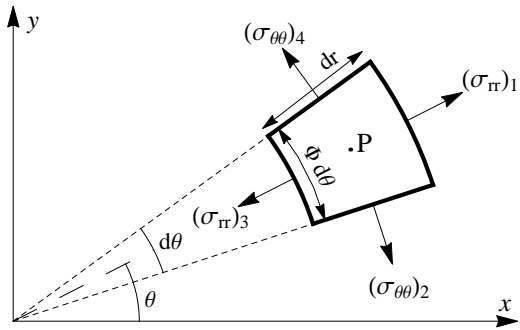} \  \  \ \ \
	\caption{Radial force balance on an infinitesimal element of a flat sheet \cite[p.~65]{timoshenko}. At the point $P$ we have contributions from the two radial stresses, $(\sigma_{rr})_1\Phi d\theta$ and $-(\sigma_{rr})_3\Phi d\theta$, and from the two azimuthal stresses $-(\sigma_{\theta\theta})_2dr\sin(d\theta/2)$ and $-(\sigma_{\theta\theta})_4dr\sin(d\theta/2)$. Balancing these terms gives  Eq.~(\ref{in-plane-eq-sigma1}). 
	}
  \label{radial-force-balance-flat}
\end{figure}

We now demonstrate the advantage of the alternative formulation in
simple examples of flat configurations. In the flat state the bending
energy is zero and the equation of equilibrium is obtained by
minimizing the stretching energy alone. To do so we first set
$\zeta=0$ in Eqs.~(\ref{epsilon-in-plane}),
\begin{subequations}\label{epsilon-planar}
\begin{eqnarray}
&&\epsilon_{rr}=\partial_r u_r, \label{epsilon-rr-flat} \\
&&\epsilon_{\theta\theta}=\frac{r}{\Phi}-1+\frac{u_r}{\Phi}, \label{epsilon-2theta-flat}
\end{eqnarray}
\end{subequations}
and then substitute Eqs.~(\ref{epsilon-planar}) in (\ref{Es-final-sigma}), obtaining,
\begin{equation}
E_s=\frac{1}{2}\int_0^R\int_0^{2\pi}\left[\sigma_{rr}\partial_r u_r +\sigma_{\theta\theta}\left(\frac{r}{\Phi}-1+\frac{u_r}{\Phi}\right)\right]\Phi d\theta dr.\label{stretching-E-flat}
\end{equation}
Minimization of $E_s$ with respect to $u_r$ gives the equation of equilibrium,
\begin{eqnarray}
&&\partial_r(\Phi\sigma_{rr})-\sigma_{\theta\theta}=0,
\label{in-plane-eq-sigma1} 
\end{eqnarray}
which expresses balance of forces in the radial direction (see Fig.~\ref{radial-force-balance-flat}).  Substituting the in-plane strains, Eqs.~(\ref{epsilon-planar}), in the stress components, Eqs.~(\ref{sigma-2D}), and then in (\ref{in-plane-eq-sigma1}), we obtain the equation of equilibrium in terms of $u_r$ alone,
\begin{eqnarray}
&&\Phi\partial_r(\Phi\partial_r u_r)-u_r=r-\Phi-\nu\Phi(1-\partial_r\Phi).
\label{eom-flat-ur} 
\end{eqnarray}
This second-order equation for $u_r$ is supplemented by two boundary conditions: vanishing stress at the free edge, $\left.\sigma_{rr}\right|_{r=R}=0$ and vanishing displacement at the origin. The resulting conditions are
\begin{subequations}\label{bc-planar}
\begin{eqnarray}
&&\left[\Phi\partial_r u_r+\nu u_r+\nu\left(r-\Phi\right)\right]_{r=R}=0, \label{eom-flat-ur-bc} \\
&&\left.u_r\right|_{r=0}=0. \label{bc-ur-zero}
\end{eqnarray}
\end{subequations}

\begin{figure}[!tbh]
\vspace{0.7cm}
  \centering
  \includegraphics[width=3.8cm]{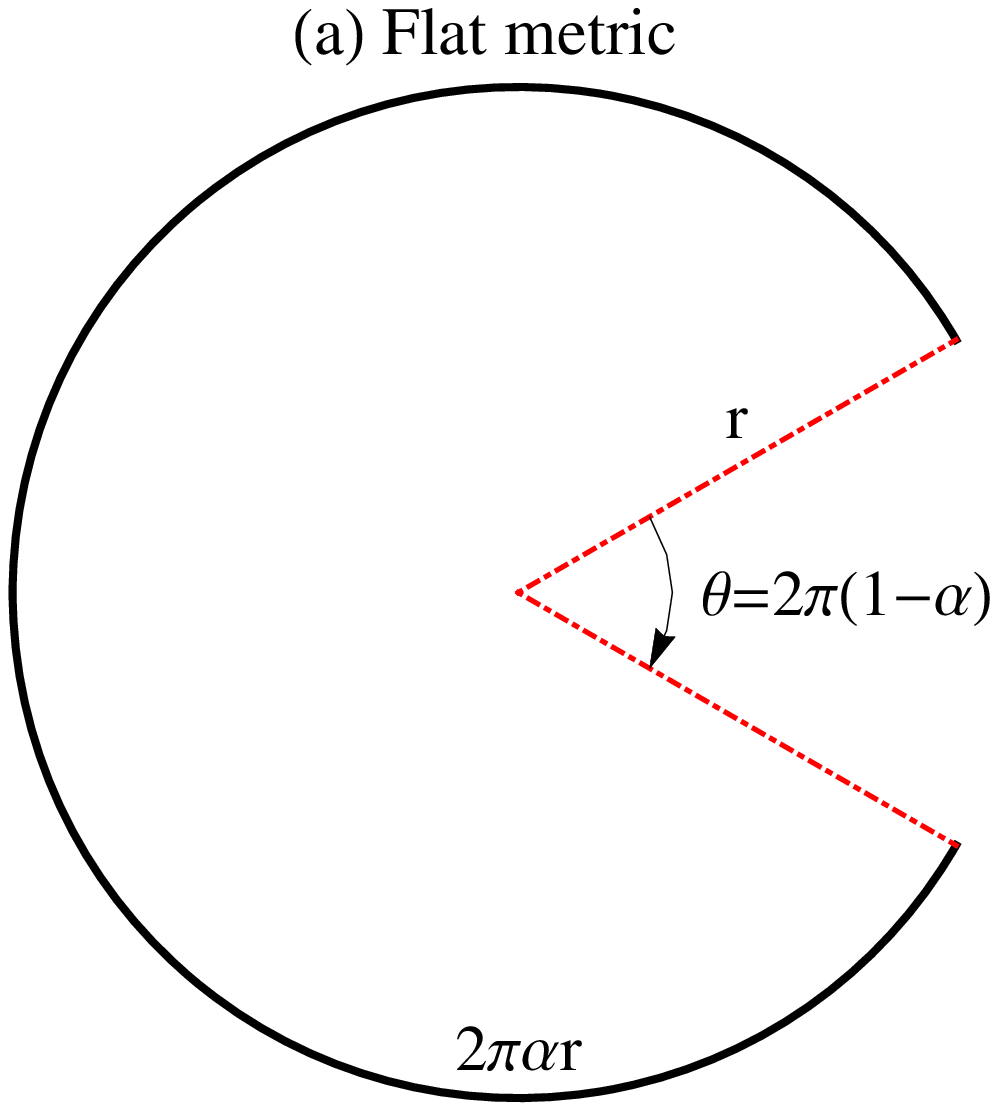} \ \ \
	 \includegraphics[width=3.5cm]{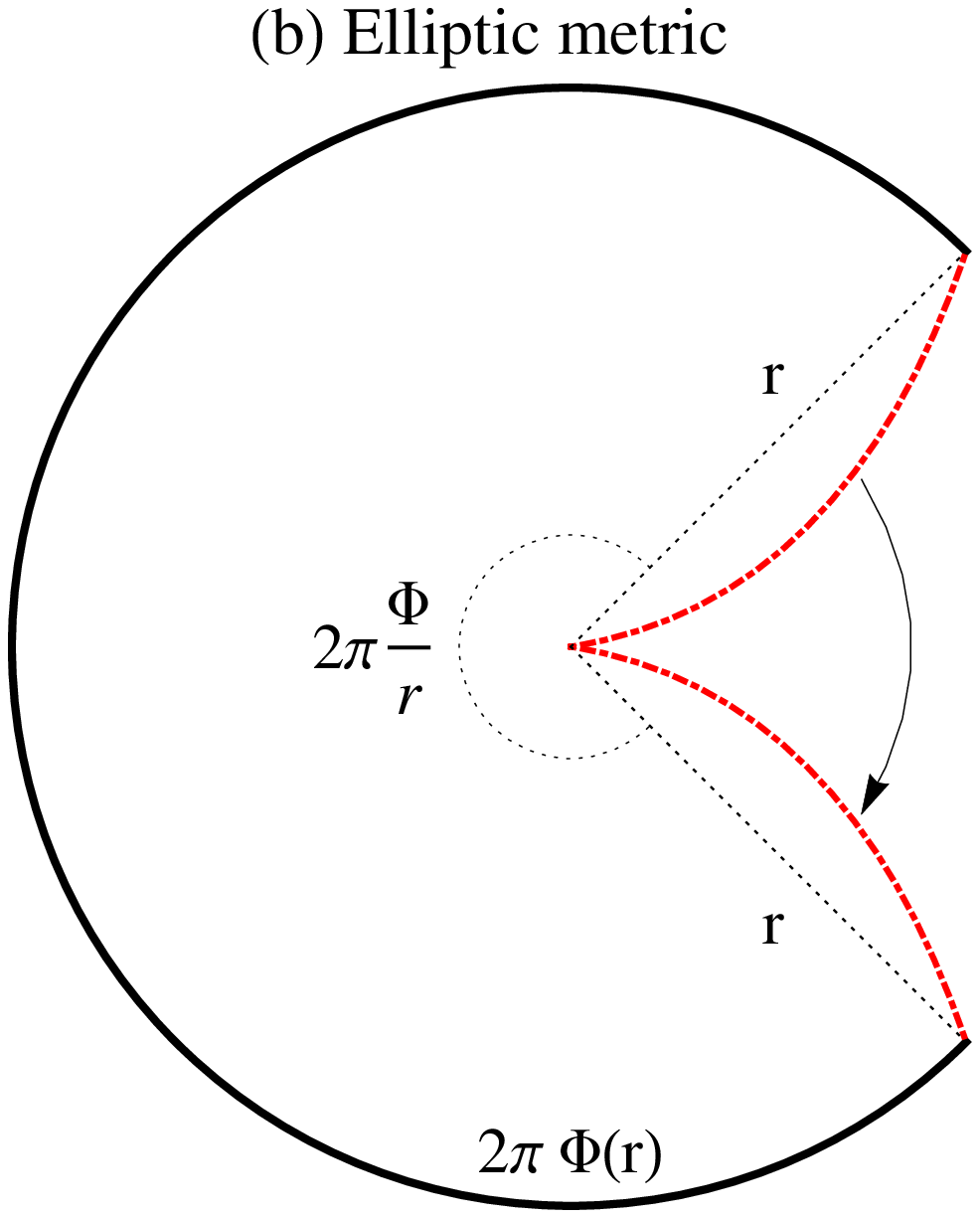} \ \ \
	 \includegraphics[width=4cm]{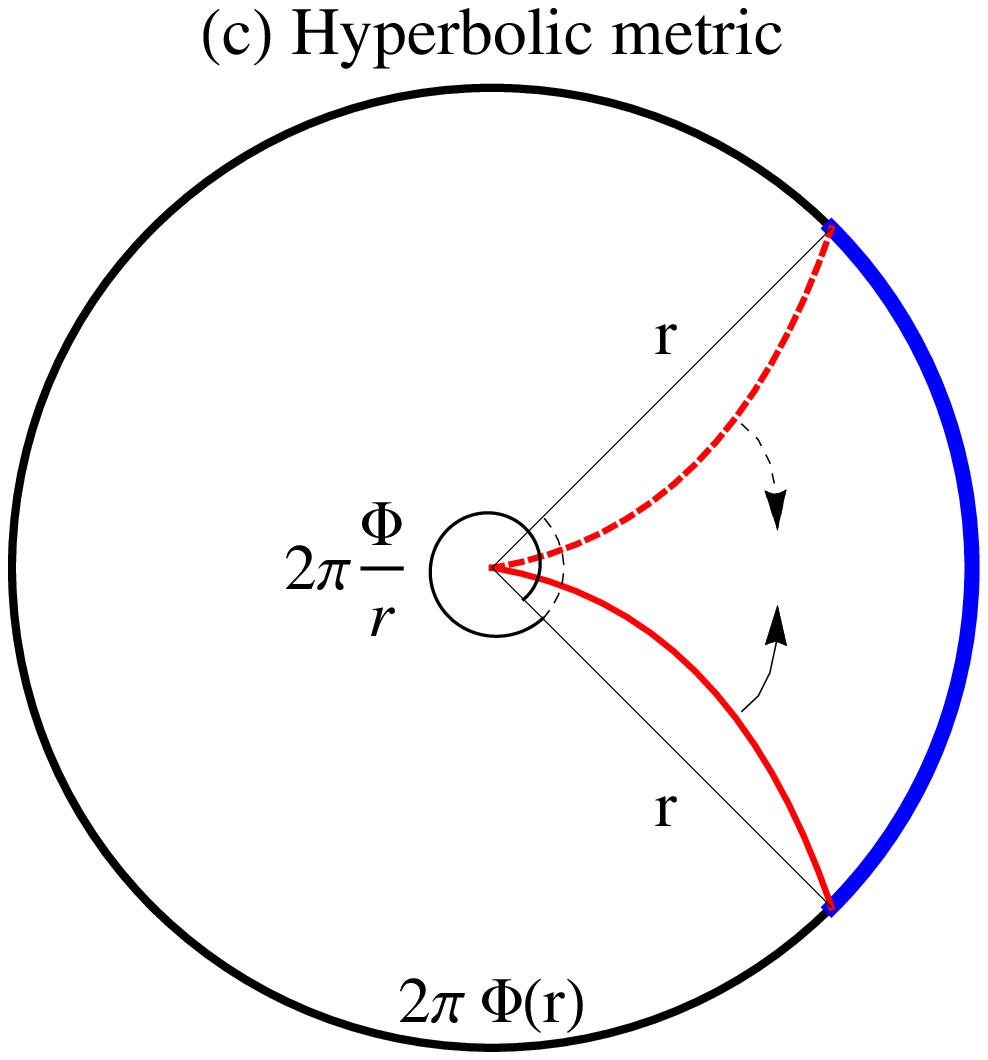}\  \  \
	\caption{Layouts of the three considered reference
          metrics. (a) Flat metric, Eq.~(\ref{Phi-flat}). When the two
          radii (dash-dotted red lines) are held together, the rest
          length of concentric circles on the closed disc become $2\pi
          \alpha r<2\pi r$. (b) Elliptic metric,
          Eq.~(\ref{Phi-elliptic}). Gluing together the two curved
          dash-dotted red lines creates a frustrated disc, where
          concentric circles have rest length of $2\pi\Phi(r)<2\pi
          r$. (c) Hyperbolic metric, Eq.~(\ref{Phi-hyperbolic}). In
          this panel dashing represents unseen lines; concentric
          circles have rest length $2\pi\Phi(r)>2\pi r$, causing
          pieces of the disc to be placed in the relaxed configuration
          one over the other (marked in blue). Attaching together the
          lower (hidden) red-dashed line with the upper solid red line
          results in a disc with a hyperbolic metric.
	}
  \label{metric-types-flat}
\end{figure}

Importantly, unlike earlier analysis of the same problem \cite{Efi2},
Eqs.~(\ref{eom-flat-ur}) and (\ref{bc-planar}) are {\it linear} and
therefore solvable. To demonstrate this key advantage we now derive
exact solutions of Eq.~(\ref{eom-flat-ur}) for three types of
reference metrics: flat, elliptic, and hyperbolic (see
Fig.~\ref{metric-types-flat}).
In the following subsections we compare the results obtained from
analytical solutions of our model for the different reference metrics
with those obtained from the ESK nonlinear equations. To assure a
meaningful comparison we examine the following: (a) the radial
displacement $u_r$, which is an unambiguous experimental observable;
(b) the stress components obtained by variation of the energy with
respect to the strain $\epsilon$ (not the metric-based one,
$\tilde\epsilon$) for {\em both} models. In the Supplemental Material
\ref{Supplementary} we elaborate on the relations between these stress tensors in
the two theories.


\subsection{Flat metric}
A flat reference  metric is given by,
\begin{equation}
\Phi(r)=\alpha r,
\label{Phi-flat}
\end{equation}
where $\alpha<1$. Substituting Eq.~(\ref{Phi-flat}) in (\ref{eom-flat-ur}) and (\ref{eom-flat-ur-bc}) gives,
\begin{subequations}\label{eom-bc-flat}
\begin{eqnarray}
&&\alpha^2 r \partial_r(r\partial_r u_r)-u_r=(1-\alpha)(1-\nu \alpha)r, \label{eom-flat-metric} \\
&&\left[\alpha r \partial_r u_r+\nu u_r+\nu(1-\alpha)r\right]_{r=R}=0. \label{bc-flat-metric}
\end{eqnarray}
\end{subequations}
Equation~(\ref{eom-flat-metric}) replaces the nonlinear Eq.~(10) of Ref.~\cite{Efi2} which could be solved only numerically. The solution to Eq.~(\ref{eom-flat-metric}) is given by,
\begin{equation}
u_r(r)=A_0 r^{1/\alpha}+B_0 r^{-1/\alpha}-\frac{1-\alpha\nu}{1+\alpha}r,
\label{ur-sol-flat}
\end{equation}
where $A_0$ and $B_0$ are constants to be determined by boundary
conditions. The vanishing displacement at the disc center,
Eq.~(\ref{bc-ur-zero}), is satisfied for $B_0=0$. The value of $A_0$ is
determined by the second boundary condition,
(\ref{bc-flat-metric}). This gives,
\begin{equation}
u_r(r)=-\frac{1-\alpha\nu}{1+\alpha}\left[1-\frac{(1-\nu)\alpha}{1-\alpha\nu}\left(\frac{r}{R}\right)^{\frac{1}{\alpha}-1}\right]r.
\label{ur-sol-flat2}
\end{equation}

\begin{figure}[!tbh]
\vspace{0.7cm}
  \centering
  \includegraphics[width=7cm]{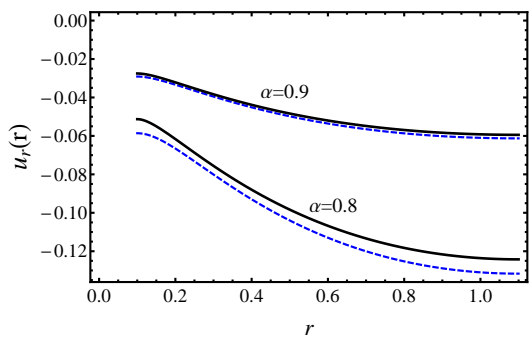} \ \ \
	\caption{Comparison between the exact solution for the radial
          displacement (Eq.~(\ref{ur-sol-flat-hole}); black, solid
          line) and the numerical solution of Eq.~(10) in
          Ref.~\cite{Efi2} (dashed, blue line) for a flat reference metric. We consider an annulus
          with inner and outer radii $R_i=0.1$ and $R_o=1.1$. In accordance
          with the example in Ref.~\cite{Efi2}, we use $\nu=0$. 
}
  \label{Flat-radial-displacement-comparison}
\end{figure}

\begin{figure}[!tbh]
\vspace{0.7cm}
  \centering
  \includegraphics[width=7cm]{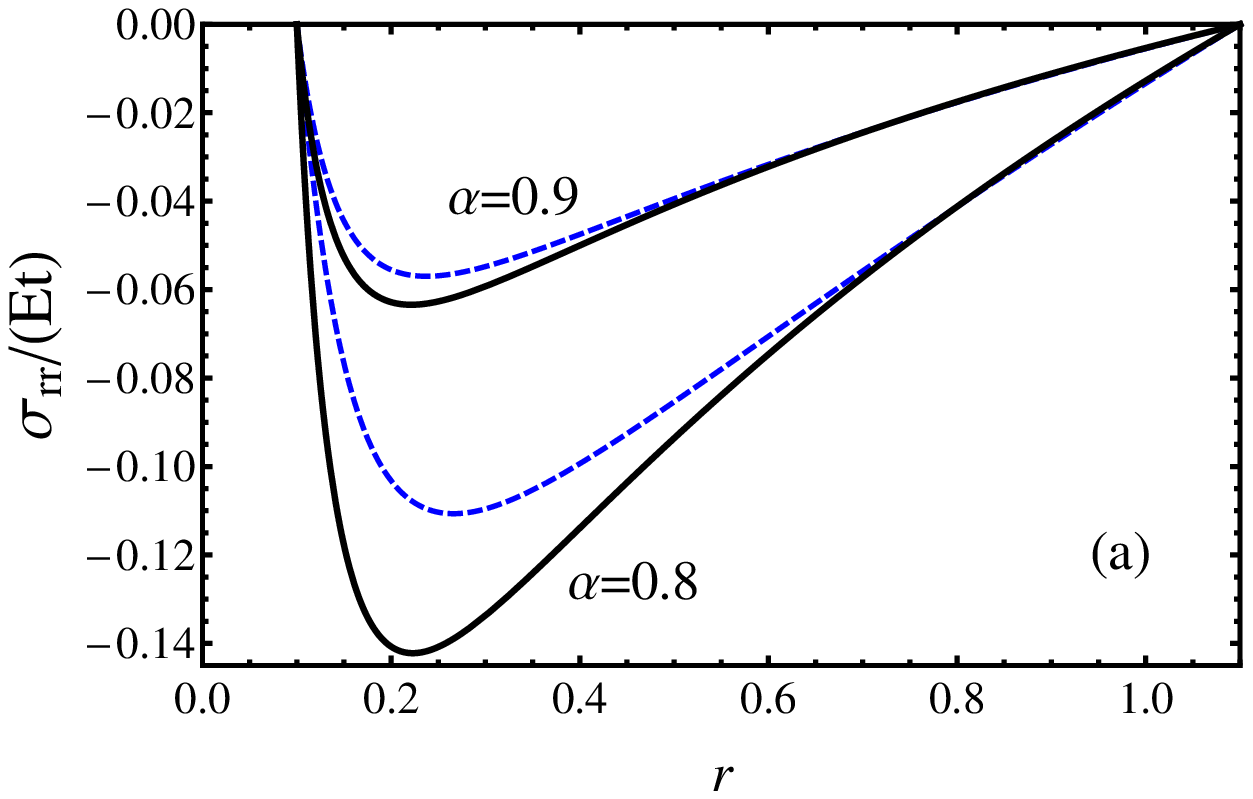} \ \ \
	\includegraphics[width=7cm]{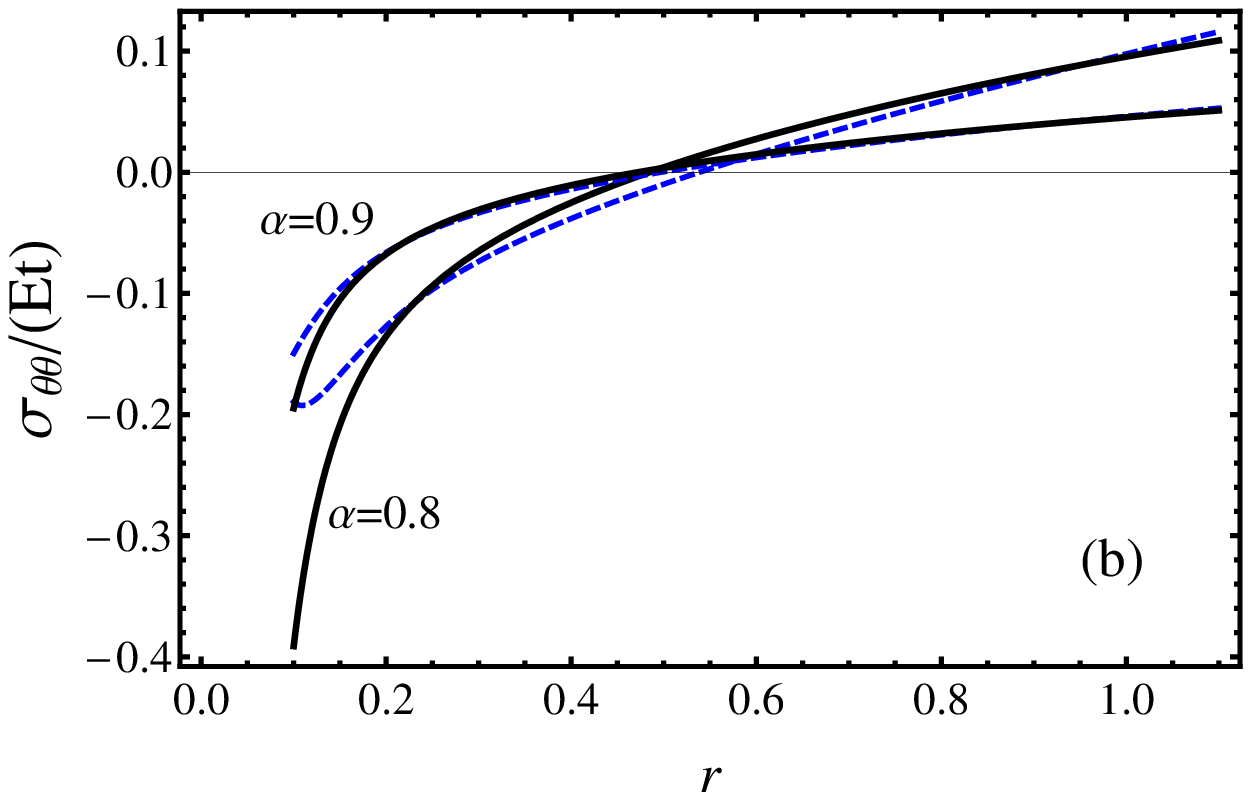}\  \  \ \ \
	\caption{Comparison between the exact plane-stress solutions
          (Eqs.~(\ref{sigma-planar-flat-sol-annulus}), black solid
          line) and the numerical solution of Eq.~(10) in
          Ref.~\cite{Efi2} (dashed blue line) for a flat reference metric. Parameters are as in
          Fig.~\ref{Flat-radial-displacement-comparison}.
}
  \label{Flat-radial-stress-comparison}
\end{figure}

Substituting Eq.~(\ref{ur-sol-flat2}) in Eqs.~(\ref{epsilon-planar})
and then in Eqs.~(\ref{sigma-2D}), we obtain the radial and azimuthal
stress components,
\begin{subequations}\label{sigma-planar-flat-sol-disc}
\begin{eqnarray}
&&\sigma_{rr}(r)=-\frac{E t}{1+\alpha}\left[1-\left(\frac{r}{R}\right)^{\frac{1}{\alpha}-1}\right], \label{sigma-rr-flat} \\
&&\sigma_{\theta\theta}(r)=-\frac{E t}{1+\alpha}\left[\alpha-\left(\frac{r}{R}\right)^{\frac{1}{\alpha}-1}\right]. \label{sigma-2theta-flat}
\end{eqnarray}
\end{subequations}
Note that the stress components do not depend on $\nu$. Note also that the azimuthal stress becomes positive at $r_{cr}=\alpha^{\alpha/(1-\alpha)}R$, whereas the radial one is always negative. The problem can be solved for other boundary conditions, e.g., for an annulus with inner radius $R_i$ and outer radius $R_o$, and with free boundary conditions at its two rims.  The solution reads,
\begin{subequations}\label{sigma-planar-flat-sol-annulus}
\begin{eqnarray}
&&u_r=\frac{\alpha(1-\nu)}{1+\alpha}\left[\frac{1-\rho^{\frac{1}{\alpha}+1}}{1-\rho^{2/\alpha}}\left(\frac{r}{R_o}\right)^{\frac{1}{\alpha}-1}-\frac{1+\nu}{1-\nu}\,\frac{1-\rho^{\frac{1}{\alpha}-1}}{1-\rho^{2/\alpha}}\left(\frac{R_i}{r}\right)^{\frac{1}{\alpha}+1}-\frac{1-\nu\alpha}{\alpha(1-\nu)}\right] r ,\label{ur-sol-flat-hole}\\
&&\sigma_{rr}=-\frac{Et}{1+\alpha}\left[1-\frac{1-\rho^{\frac{1}{\alpha}+1}}{1-\rho^{2/\alpha}}\left(\frac{r}{R_o}\right)^{\frac{1}{\alpha}-1}-\frac{1-\rho^{\frac{1}{\alpha}-1}}{1-\rho^{2/\alpha}}\left(\frac{R_i}{r}\right)^{\frac{1}{\alpha}+1}\right], \label{sigmarr-sol-flat-annulus}
  \\ &&\sigma_{\theta\theta}=-\frac{Et}{1+\alpha}\left[\alpha-\frac{1-\rho^{\frac{1}{\alpha}+1}}{1-\rho^{2/\alpha}}\left(\frac{r}{R_o}\right)^{\frac{1}{\alpha}-1}+\frac{1-\rho^{\frac{1}{\alpha}-1}}{1-\rho^{2/\alpha}}\left(\frac{R_i}{r}\right)^{\frac{1}{\alpha}+1}\right] \label{sigma2theta-sol-flat-annulus},
\end{eqnarray}
\end{subequations}
where $\rho\equiv R_i/R_o$.  In
Fig.~\ref{Flat-radial-displacement-comparison} we compare the exact
analytical solution for the radial displacement, Eq.~(\ref{ur-sol-flat-hole}), with the numerical
solution of the formalism given in Ref.~\cite{Efi2}. The two theories
converge to the same solution as $\alpha\rightarrow 1$. However, away
from this nearly Euclidean regime there are significant differences in
the resultant displacements. Since the displacement is an unambiguous
observable, these differences underline the fact that the two
formulations are not equivalent. Figure
\ref{Flat-radial-stress-comparison} presents a similar comparison of
the plane stresses obtained from the two theories.

\subsection{Elliptic metric}
\label{Elliptic-metric-flat}
An elliptic reference metric is given by,
\begin{equation}
\Phi(r)=\frac{1}{\sqrt{K}}\sin(\sqrt{K}r),
\label{Phi-elliptic}
\end{equation}
where $K$ is a constant positive reference Gaussian curvature.
Substituting Eq.~(\ref{Phi-elliptic}) in Eqs.~(\ref{eom-flat-ur})  and (\ref{eom-flat-ur-bc}) gives,
\begin{subequations}\label{elliptic-eom-bc}
\begin{eqnarray}
&&\sin (r)\partial_r(\sin (r)\partial_r u_r)-u_r=r-\sin (r) -\nu\sin (r)(1-\cos (r)), \label{eom-elliptic-flat} \\
&&\left[\sin (r)\partial_r u_r+\nu u_r+\nu(r-\sin (r))\right]_{r=R}=0, \label{bc-elliptic-flat}
\end{eqnarray}
\end{subequations}
where we have rescaled the lengths $r$ and $u_r$ by $K^{-1/2}$. The following expression is verified to be the general solution by direct substitution in Eq.~(\ref{eom-elliptic-flat}),
\begin{equation}
u_r(r)=A_0 \tan(r/2)+B_0 \cot(r/2)-r-2(1+\nu)\cot(r/2)\ln[\cos(r/2)]. \label{ur-elliptic-sol}
\end{equation}
We set $B_0=0$  to satisfy the vanishing displacement at the disc center, Eq.~(\ref{bc-ur-zero}), and determine $A_0$ by the boundary condition (\ref{bc-elliptic-flat}), obtaining,
\begin{eqnarray}
u_r(r)=-r-2(1-\nu)\ln[\cos(R/2)]\cot^2(R/2)\left(1+\frac{1+\nu}{1-\nu}\frac{\cot^2(r/2)}{\cot^2(R/2)}\frac{\ln[\cos(r/2)]}{\ln[\cos(R/2)]}\right)\tan(r/2). \label{ur-elliptic-sol2}
\end{eqnarray}
Note that the solution diverges for $r=r_n=n\pi$ where $n$ is a
positive integer. At such points the reference metric,
Eq.~(\ref{Phi-elliptic}), vanishes, i.e., these divegencies correspond
to unphysical cases where the rest length shrinks to
zero. Substituting Eq.~(\ref{ur-elliptic-sol2}) in
Eqs.~(\ref{sigma-2D}), we obtain the distributed stress on the disc,
\begin{subequations}\label{sigma-disc-elliptic}
\begin{eqnarray}
&&\sigma_{rr}(r)=-Et\left(1-\frac{\cot^2(r/2)}{\cot^2(R/2)}\frac{\ln[\cos(r/2)]}{\ln[\cos(R/2)]}\right)\frac{\ln[\cos(R/2)]\cot^2(R/2)}{\cos^2(r/2)}, \label{sigma-rr-Ellitic-sol} \\
&&\sigma_{\theta\theta}(r)=-Et\left(1+\frac{\ln[\cos(r/2)]}{\sin^2(r/2)}+\cot^2(R/2)\frac{\ln[\cos(R/2)]}{\cos^2(r/2)}\right).  \label{sigma-2theta-Ellitic-sol} 
\end{eqnarray}
\end{subequations}
Once again, the solution is independent of the Poisson ratio.

\begin{figure}[!tbh]
\vspace{0.7cm}
  \centering
  \includegraphics[width=7cm]{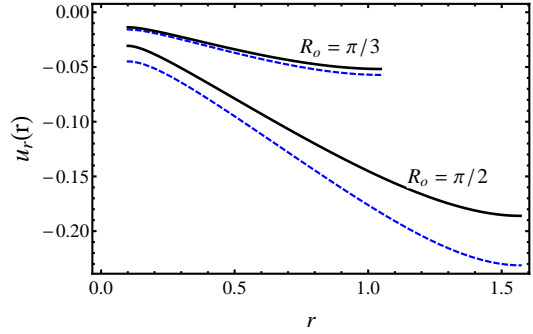} \ \ \
	\caption{The exact solution for the radial displacement
          (Eqs.~(\ref{ur-elliptic-sol}) and
          (\ref{ur-sol-Elliptic-hole}); black solid line) is plotted
          alongside the numerical solution of Eq.~(10) in
          Ref.~\cite{Efi2} (dashed blue line) for an elliptic reference metric. We consider an annulus
          with a normalized inner radius $R_i=0.1$, $\nu=0$, and two different values of
          $R_o$ as indicated.
}
  \label{Elliptic-radial-displacement-comparison}
\end{figure}

\begin{figure}[!tbh]
\vspace{0.7cm}
  \centering
  \includegraphics[width=7cm]{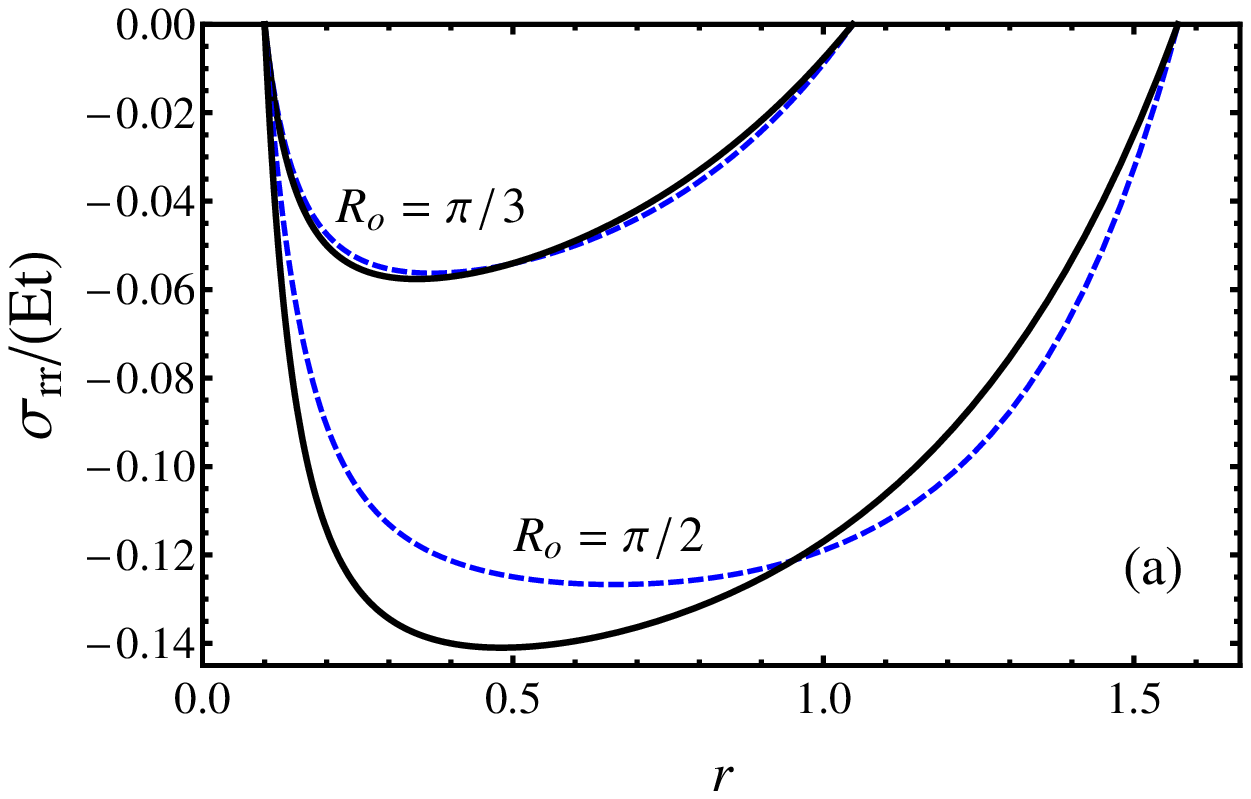} \ \ \ 
	\includegraphics[width=7cm]{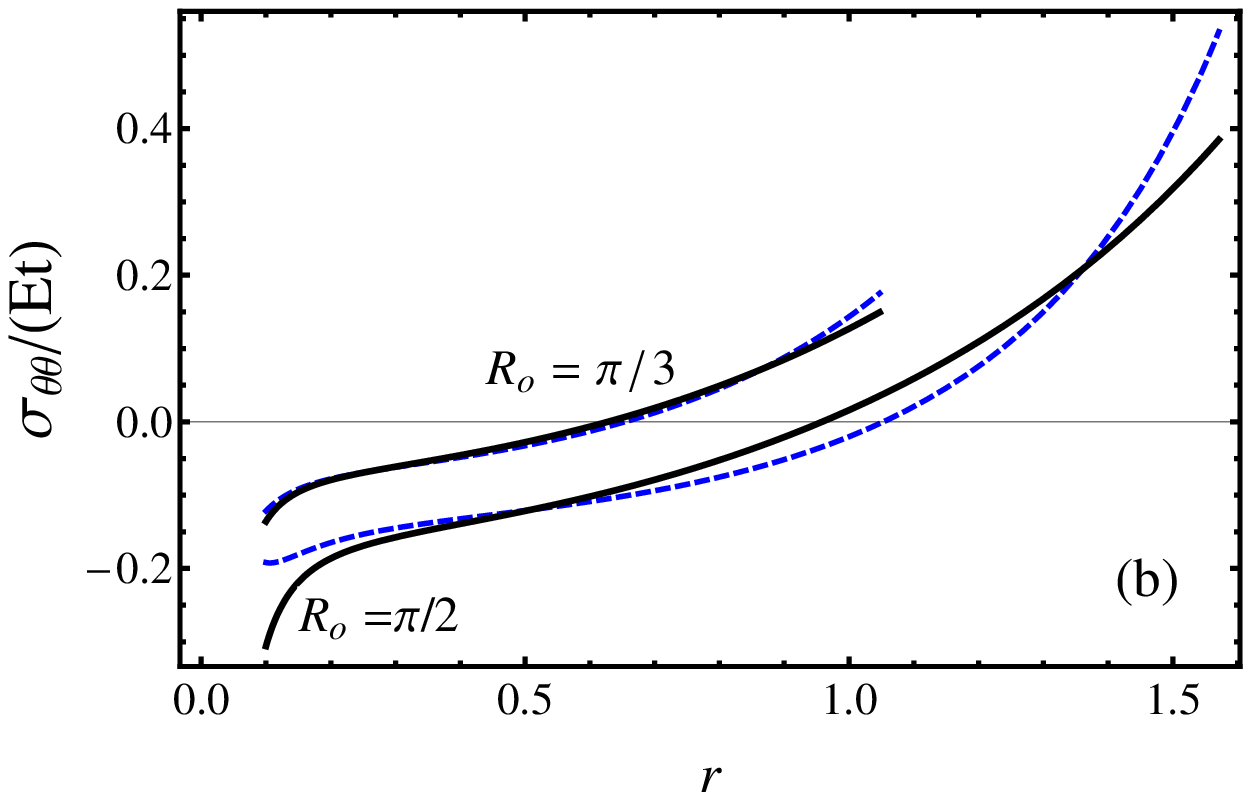}\  \  \ \ \
	\caption{Comparison between the exact plane-stress solutions,
          Eqs.~(\ref{sigma-annulus-elliptic}) (black solid line), and
          the numerical results based on Ref.~\cite{Efi2} (dashed blue
          line) for an elliptic reference metric. Parameters as in
          Fig.~\ref{Elliptic-radial-displacement-comparison}.
}
  \label{Elliptic-radial-stress-comparison}
\end{figure}

In order to compare our exact solution to the numerical one obtained
in Ref.~\cite{Efi2}, we also derive the displacement and the planar
stress in an annulus with free boundary conditions. In this case the
constants $A_0$ and $B_0$ in Eq.~(\ref{ur-elliptic-sol}) are
\begin{subequations}\label{ur-sol-Elliptic-hole}
\begin{eqnarray}
&&A_0=\frac{4(1-\nu)}{\cos(R_i)-\cos(R_o) }\cos^2(R_i/2)\cos^2(R_o/2)\left(\ln[\cos(R_i/2)]-\ln[\cos(R_o/2)]\right),\\
&&B_0=\frac{1+\nu}{\cos(R_i)-\cos(R_o)}\left[(1+\cos(R_i))(1-\cos(R_o))\ln[\cos(R_i/2)]-(1-\cos (R_i))(1+\cos(R_o))\ln[\cos(R_o/2)]\right],\nonumber \\
\end{eqnarray}
\end{subequations}
and the stress components become 
\begin{subequations}\label{sigma-annulus-elliptic}
\begin{eqnarray}
\sigma_{rr}&=&-2Et\left[1+\left(1-\frac{\cos(R_i)-\cos(R_o)}{\cos(r)-\cos(R_o)}\frac{\cos^2(r/2)}{\cos^2(R_i/2)}\frac{\ln[\cos(r/2)]}{\ln[\cos(R_i/2)]}\right)\frac{1+\cos(R_i)}{1+\cos(R_o)}\frac{\cos(r)-\cos(R_o)}{\cos(R_i)-\cos(r)}\frac{\ln[\cos(R_i/2)]}{\ln[\cos(R_o/2)]}\right] \nonumber \\
&\times& \frac{1+\cos(R_o)}{\sin^2(r)}\frac{\cos(R_i)-\cos(r)}{\cos(R_i)-\cos(R_o)}\ln[\cos(R_o/2)], \label{sigmarr-sol-elliptic-annulus} \\
\sigma_{\theta\theta}&=&-Et\left[1+\frac{\ln[\cos(r/2)]}{\sin^2(r/2)}+4\left(1-\frac{\cos^2(R_i/2)}{\cos^2(R_o/2)}\frac{1-\cos(R_o) \cos(r)}{1-\cos(R_i)\cos(r)}\frac{\ln[\cos(R_i/2)]}{\ln[\cos(R_o/2)]}\right)\right. \nonumber \\
&\times&\left. \frac{\cos^2(R_o/2)}{\sin^2(r)}\frac{1-\cos(R_i)\cos(r)}{\cos(R_i)-\cos(R_o)}\ln[\cos(R_o/2)]\right]. \label{sigma2theta-sol-elliptic-annulus}
\end{eqnarray}  
\end{subequations}
In Fig.~\ref{Elliptic-radial-displacement-comparison} we compare the
radial displacement obtained from this exact solution,
Eqs.~(\ref{ur-elliptic-sol}) and (\ref{ur-sol-Elliptic-hole}), to the
numerical solution of Eq.~(10) in Ref.~\cite{Efi2}. In addition,
Fig.~\ref{Elliptic-radial-stress-comparison} compares the radial and
azimuthal stress components of the two models. The two solutions
converge for a narrow annulus and differ significantly as the annulus
becomes wider. (Note that increasing $R_o$ is equivalent to increasing
$K$.)

\subsection{Hyperbolic metric}
A hyperbolic reference metric is given by,
\begin{equation}
\Phi(r)=\frac{1}{\sqrt{K}}\sinh(\sqrt{K}r).
\label{Phi-hyperbolic}
\end{equation}
The equation of equilibrium and the boundary condition are obtained by substituting Eq.~(\ref{Phi-hyperbolic}) in Eq.~(\ref{eom-flat-ur}) and (\ref{eom-flat-ur-bc}),
\begin{subequations}\label{eom-hyperbolic-planar}
\begin{eqnarray}
&&\sinh (r)\partial_r(\sinh (r)\partial_r u_r)-u_r=r-\sinh (r)+\nu\sinh (r)(1-\cosh (r)), \label{eom-hyperbolic-flat} \\
&&\left[\sinh (r)\partial_r u_r+\nu u_r+\nu(r-\sinh(r))=0\right]_{r=R}, \label{bc-hyperbolic-flat}
\end{eqnarray}
\end{subequations}
where again we have rescaled $r$ and $u_r$ by $K^{-1/2}$. Since Eq.~(\ref{eom-hyperbolic-flat}) is obtained from (\ref{eom-elliptic-flat}) by a Wick transformation,
\begin{eqnarray}
&&r\rightarrow i r,  \ \ \ \  u_r\rightarrow i u_r,
\label{WicksHyperbolic}
\end{eqnarray} 
we immediately obtain from Eqs.~(\ref{ur-elliptic-sol2}) and (\ref{sigma-disc-elliptic}) the solution, 
\begin{subequations}\label{ur-sigma-disc-hyperbolic}
\begin{eqnarray}
&&u_r(r)=-r+2(1-\nu)\coth^2(R/2)\ln[\cosh(R/2)]\left(1+\frac{1+\nu}{1-\nu}\,\frac{\coth^2(r/2)}{\coth^2(R/2)}\frac{\ln[\cosh(r/2)]}{\ln[\cosh(R/2)]}\right)\tanh(r/2), \label{ur-hyperbolic-sol-flat} \\
&&\sigma_{rr}(r)=Et\left(1-\frac{\coth^2(r/2)}{\coth^2(R/2)}\frac{\ln[\cosh(r/2)]}{\ln[\cosh(R/2)]}\right)\frac{\ln[\cosh(R/2)]\coth^2(R/2)}{\cosh^2(r/2)}, \label{sigma-rr-hyperbolic-flat} \\
&&\sigma_{\theta\theta}(r)=-Et\left(1-\frac{\ln[\cosh(r/2)]}{\sinh^2(r/2)}-\cosh^2(R/2)\frac{\ln[\cosh(R/2)]}{\cosh^2(r/2)}\right). \label{sigma-2theta-hyperbolic-flat}
\end{eqnarray}
\end{subequations}
It is readily verified that this solution satisfies the boundary condition (\ref{bc-hyperbolic-flat}).

\begin{figure}[!tbh]
\vspace{0.7cm}
  \centering
  \includegraphics[width=7cm]{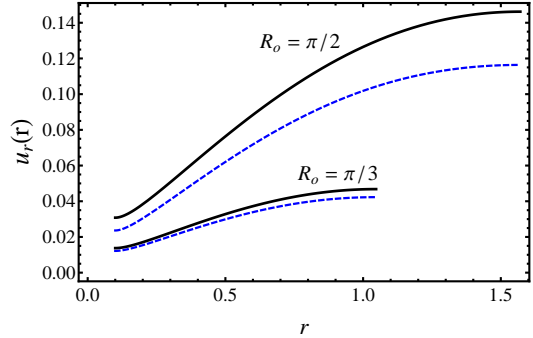} \ \ \
	\caption{The exact solution for the radial displacement (black
          solid line) is plotted alongside the numerical solution of
          Eq.~(10) in Ref.~\cite{Efi2} (dashed blue line) for a hyperbolic reference metric. We consider
          an annulus with  inner normalized radius
          $R_i=0.1$, $\nu=0$, and two different values of $R_o$ as
          indicated. 
}
  \label{Hyperbolic-radial-displacement-comparison}
\end{figure}

\begin{figure}[!tbh]
\vspace{0.7cm}
  \centering
  \includegraphics[width=7cm]{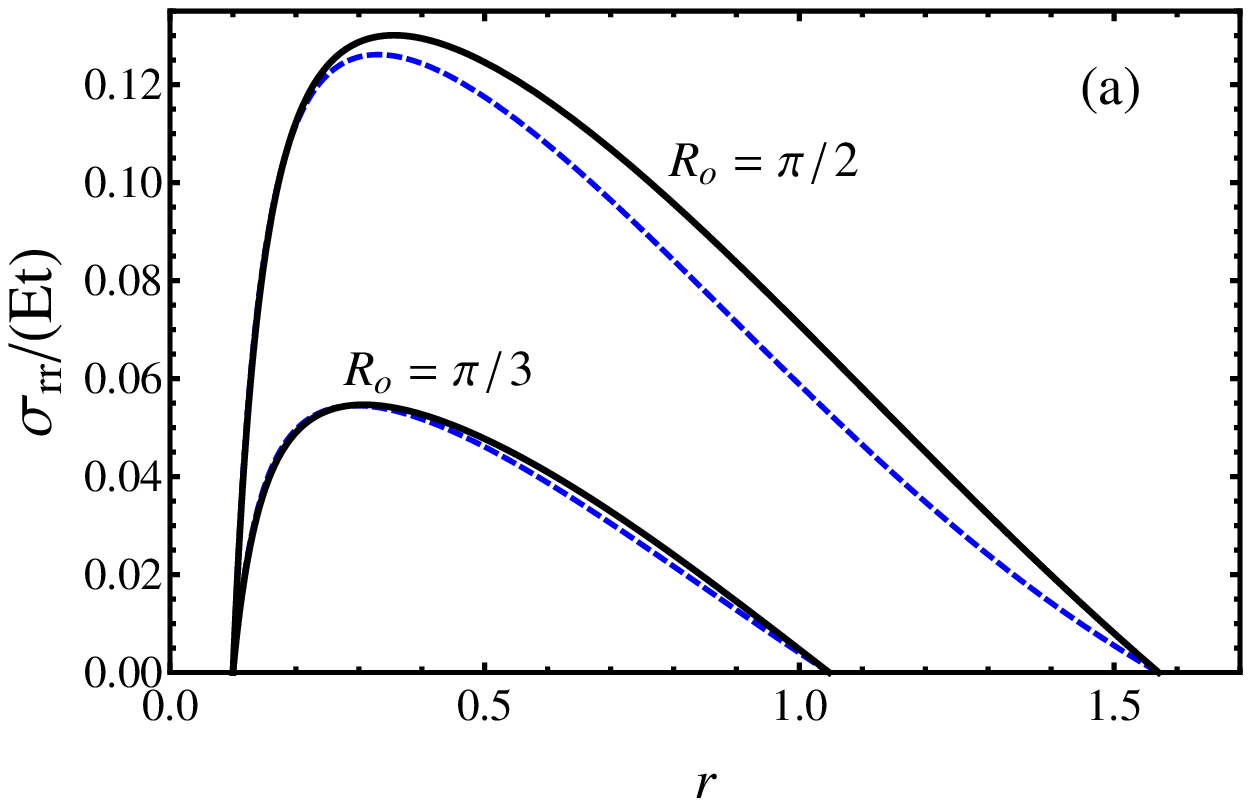} \ \ \
	\includegraphics[width=7cm]{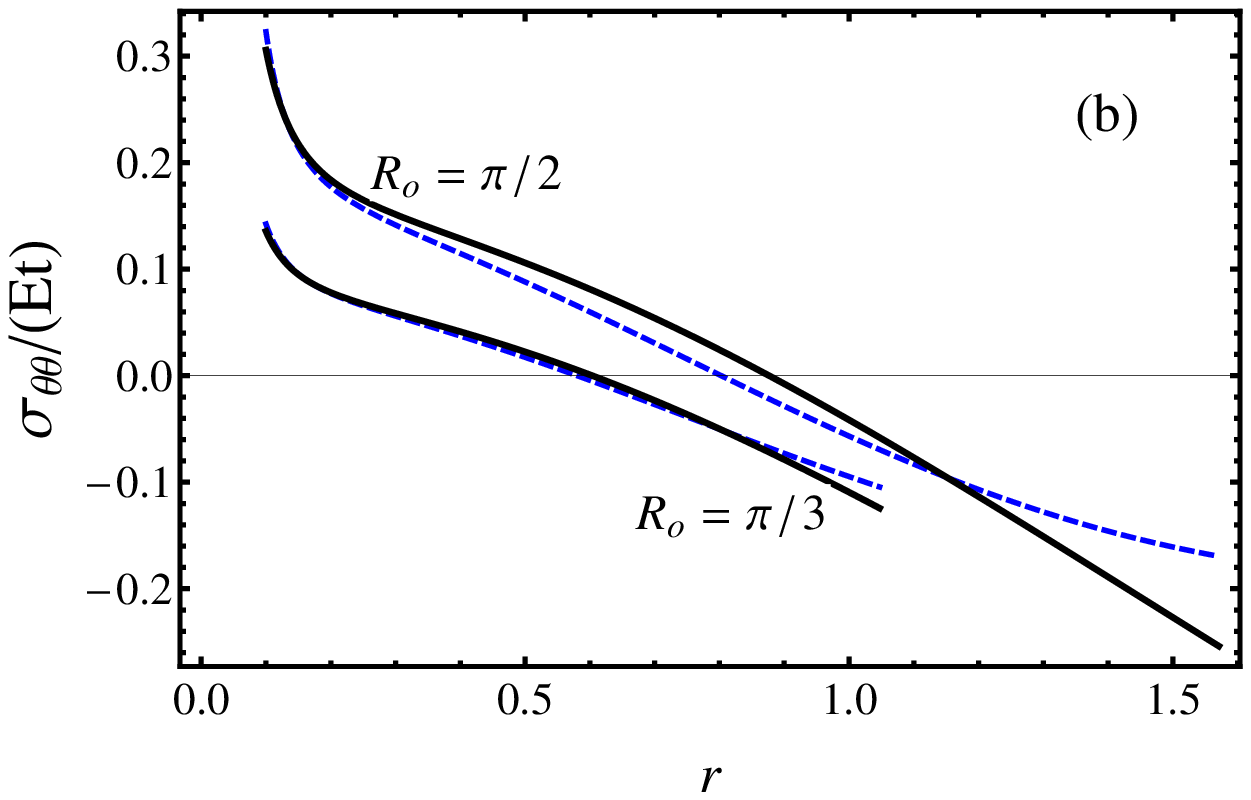}\  \  \ \ \
	\caption{The exact radial and azimuthal plane-stress solutions
          for a flat annulus with a hyperbolic reference metric (solid
          black line) are compared with the numerical solution of
          Eq.~(10) in Ref.~\cite{Efi2} (dashed blue line). Parameters
          are as in
          Fig.~\ref{Hyperbolic-radial-displacement-comparison}.
}
  \label{Hyperbolic-radial-stress-comparison}
\end{figure}

Similarly, the radial displacement and the stress distribution in an annulus with hyperbolic reference metric is obtained from Eqs.~(\ref{sigma-annulus-elliptic}) via a Wick transformation, Eq.~(\ref{WicksHyperbolic}). In Figs.~\ref{Hyperbolic-radial-displacement-comparison} and~\ref{Hyperbolic-radial-stress-comparison} we compare these solutions to the one obtained in Ref.~\cite{Efi2}.

\section{Stability criterion for isometric immersions}
\label{self-consis}

An isometric immersion refers to a strain-free configuration,
$\epsilon_{\alpha\beta}\equiv 0$, leading to $E_s=0$. It is obviously
the minimizer of the elastic energy for $B=0$. In this section we do
not directly seek the minimizer of the total energy,
Eq.~(\ref{E-2D-final}), but check whether the isometric immersion
happens to be a minimizer also for $B>0$.  Since this configuration
already minimizes $E_s$, we need to check only whether it also
minimizes $E_b$. Note, however, that there are two different routes
for such minimization: (a) set $\epsilon_{\alpha\beta}=0$ in $E_b$ and then
minimize with respect to curvature alone; (b) minimize $E_b$
with respect to both strain and curvature and only then set the strain
to zero, which is the appropriate route. It is straightforward to show that in our model the two
routes are equivalent. This is because the strain appears only
quadratically in the energy (see, for example, Eq.~(\ref{E-1D-no-P}))
and, therefore, setting the strain to zero, either before or after
minimization, eliminates the same terms. However, in the ESK model the
additional coupling term in the bending energy is linear in the strain
(compare, for example, to Eq.~(\ref{ESK-reduced})), leading to
different results of the two routes. Hence, we conclude that the two
theories should give the same results in case (a) but may differ in the appropriate minimization, 
case (b).

For a given reference metric of the form of Eq.~(\ref{ref-metric}),
i.e., for a given $\Phi(r)$, the requirement of vanishing strain
uniquely determines the configuration of the sheet up to rigid
transformations. Indeed, setting Eqs.~(\ref{epsilon-in-plane}) to
zero, we obtain,
\begin{subequations}\label{iso-displacemets}
\begin{eqnarray}
&&u_r(r)=\Phi-r, \label{ur-isometric}\\
&&\partial_r\zeta=\sqrt{1-(\partial_r\Phi)^2}. \label{zeta-isometric}
\end{eqnarray}
\end{subequations}
We can now check whether this configuration satisfies local mechanical equilibrium of bending moments. 

We substitute in Eq.~(\ref{Eb-final-M})  $\phi_{rr}=\partial_r\phi^r$ and $\phi_{\theta\theta}=\sin\phi^r/\Phi$ (see Fig.~\ref{epsilon-rr-2theta}(a)),  
\begin{equation}
E_b=\frac{1}{2}\int_0^R\int_0^{2\pi}\left[ M_{rr}\partial_r\phi^r+M_{\theta\theta}\sin\phi^r/\Phi\right]\Phi d\theta dr,
\label{Eb-isometry}
\end{equation} 
and minimize with respect to $\phi^r$,
\begin{subequations}\label{moment-balance-eom-bc}
\begin{eqnarray}
&&\partial_r(\Phi M_{rr})-\cos\phi^r M_{\theta\theta}=0, \label{moment-balance-eom} \\
&&\left.M_{rr}\right|_{r=R}=0. \label{moment-balance-bc}
\end{eqnarray}
\end{subequations}
(As has been done for the uniaxial bending case
(Appendix~\ref{app-1D-minimization}), one can show here as well that
this minimization is equivalent to the appropriate one with respect to
the spatial configuration; see Supplemental Material \ref{Supplementary}.)
Equation (\ref{moment-balance-eom}) expresses balance of moments on an
infinitesimal sheet element in the radial direction
\cite{Libai,Libai2}. The boundary condition,
Eq.~(\ref{moment-balance-bc}), imposes the vanishing of radial bending
moment at the free edge.
 
Our aim now is to check whether the displacements given by Eqs.~(\ref{iso-displacemets}) also satisfy Eqs.~(\ref{moment-balance-eom-bc}). To this end we first express $\phi_{rr}$ and $\phi_{\theta\theta}$ in terms of $\Phi(r)$ using Eqs.~(\ref{bending-strains}) and (\ref{iso-displacemets}),
\begin{subequations}\label{phi-isometry}
\begin{eqnarray}
&&\phi_{rr}=-\partial_{rr}\Phi/\sqrt{1-(\partial_r\Phi)^2}, \label{phi-rr-isometry} \\
&&\phi_{\theta\theta}=\sqrt{1-(\partial_r\Phi)^2}/\Phi. \label{phi-2theta-isometry}
\end{eqnarray}
\end{subequations}
In addition, we have (see the relation between $\phi_{\theta\theta}$ and the angle $\phi^r$ in Fig.~\ref{epsilon-rr-2theta}(a) and its caption),
\begin{equation}
\cos\phi^r=\partial_r\Phi.
\label{cos-phir-Phi}
\end{equation}
Substituting Eqs.~(\ref{phi-isometry}) in Eqs.~(\ref{M-2D}), and the result, along with Eq.~(\ref{cos-phir-Phi}),  in Eqs.~(\ref{moment-balance-eom-bc}),  we obtain an equation and a boundary condition for $\Phi(r)$ alone,
\begin{subequations}\label{sel-cosis-eom-bc}
\begin{eqnarray}
&&\partial_r(\Phi \partial_{rr}\Phi/\sqrt{1-(\partial_r\Phi)^2})+\partial_r\Phi\sqrt{1-(\partial_r\Phi)^2}/\Phi=0,
\label{Phi-self-c-eom} \\
&&\left[\partial_{rr}\Phi/\sqrt{1-(\partial_r\Phi)^2}-\nu \sqrt{1-(\partial_r\Phi)^2}/\Phi\right]_{r=R}=0. \label{Phi-self-c-bc}
\end{eqnarray}
\end{subequations}
Equations (\ref{sel-cosis-eom-bc}) are a self-consistency condition
for the reference metric, which must be satisfied for the isometric
immersion to be an equilibrium configuration of the total energy. (It
should be stressed that, if a certain isometric immersion does not
satisfy this condition, it can still become the equilibrium
configuration asymptotically, in the limit of vanishing $B/Y$
\cite{refId0}.)

The displacements, Eqs.~(\ref{iso-displacemets}), and the bulk
equilibrium equation, Eq.~(\ref{Phi-self-c-eom}), do not depend on
$\nu$. Hence, any solution but the trivial flat configuration,
$\Phi(r)=r$, will violate, in general, the boundary condition
(\ref{Phi-self-c-bc}), which does depend on $\nu$ explicitly. In
Ref.~\cite{Efi2} it was shown that such boundary conditions may be
taken care of by boundary layers.  Thus, up to a small correction at
the boundary (which vanishes in the limit of zero thickness), an
isometry that satisfies the bulk condition,
Eq.~(\ref{Phi-self-c-eom}), may be in mechanical equilibrium even if
the boundary condition (\ref{Phi-self-c-bc}) is not satisfied.

Let us now check the stability condition, Eq.~(\ref{Phi-self-c-eom}),
for the examples of flat and elliptic reference metrics.  In the case
of a hyperbolic one, Eq.~(\ref{Phi-hyperbolic}), the isometric
immersion is not a surface of revolution \cite{klein}, and therefore
lies outside the scope of this work.  (Substituting
Eq.~(\ref{Phi-hyperbolic}) in the height function,
Eq.~(\ref{zeta-isometric}), produces an imaginary result.)

Considering a flat reference metric, $\Phi(r)=\alpha r$, we
immediately find that the self-consistency condition,
Eq.~(\ref{Phi-self-c-eom}), is violated, and conclude that any
isometric immersion of this metric will be unstable for $B>0$.  The
isometric immersion of the flat metric is a cone with an opening angle
$\mathcal{\vartheta}=2\tan^{-1}(\alpha/\sqrt{1-\alpha^2})$,
\begin{equation}
{\bf f}(r,\theta)=r\left[\alpha {\bf\hat{r }}+\sqrt{1-\alpha^2}{\bf \hat{z}}\right].
\end{equation}
Note again that this does not preclude the possibility that the actual
minimizer approaches a cone asymptotically for a vanishingly small
$B/Y$ \cite{refId0}.


In the example of an elliptic reference metric we substitute
Eq.~(\ref{Phi-elliptic}) in (\ref{Phi-self-c-eom}) and find that the
self-consistency condition is satisfied in the bulk.  The isometric
immersion of an elliptic reference metric is a spherical cap of radius
$1/\sqrt{K}$,
\begin{equation}
{\bf f}(r,\theta)=\frac{1}{\sqrt{K}}\left(\sin(\sqrt{K}r)\hat{{\bf r}}+\cos(\sqrt{K}r)\hat{{\bf z}}\right).
\label{f-shape-elliptic}
\end{equation} 
When we substitute this configuration in the formalism of
Ref.~\cite{Efi1} (the first of Eqs.~(3.10) in Ref.~\cite{Efi1}), we
find that it does not satisfy balance of normal forces
(see Supplemental Material \ref{Supplementary}). This procedure corresponds to route (b) described above,
i.e., substitution of the isometric immersion in the full equations of
equilibrium rather than eliminating the strain from the beginning.
Thus, as anticipated above, the two theories disagree. A spherical cap
satisfies our stability condition but is found to be unstable for $B>0$ by the
ESK theory. (Recall that the two theories do coincide if one wrongly
follows the other route in the ESK model.)  The spherical cap
configuration of a sheet with elliptic reference metric was found to
be stable in experiments \cite{klein}.  We note that the criterion at
the boundary, Eq.~(\ref{Phi-self-c-bc}), is not satisfied by the
elliptic $\Phi(r)$. This can be mended by a thin boundary layer of
width $\propto t^{1/2}$ \cite{Efi2}. In
Appendix~\ref{boundary-layer-elliptic} we give an alternative, more
complete derivation of this result within the FvK approximation.

In Appendix~\ref{negative-Gauss-curv} we add a similar stability
criterion for two examples of surfaces of revolution whose reference
metric is slightly more general than the ones assumed so far.

\section{Discussion}
\label{conclusions}

We have presented an alternative formulation for the elasticity of
incompatible thin sheets, which is restricted to axisymmetric
deformations. This formulation and the existing ESK theory \cite{Efi2}
are not equivalent. The lack of equivalence has been demonstrated in
three systems\,---\,the existence or absence of in-plane strain in a
uniaxially bent sheet (Sec.~\ref{cylindrical-symmetry}); the strains
forming in flat incompatible sheets (Sec.~\ref{exact-sol-planar}, see
Figs.~\ref{Flat-radial-displacement-comparison} and
\ref{Elliptic-radial-displacement-comparison}); and the stability of
the spherical-cap isometry for a sheet with an elliptic reference
metric (Sec.~\ref{self-consis}).

The key ingredient that sets the two models apart is a coupling
between stretching and bending, which appears in the ESK model upon
dimensional reduction, and is removed in the present formulation by
using distance deviations, rather than metric deviations, to define
strain. (Recall, for example, Eq.~(\ref{E-1D-no-P})
vs.\ Eq.~(\ref{ESK-reduced}).) Let us pinpoint the stage at which
this difference emerges. If the derivation of
Eqs.~(\ref{dfstar})--(\ref{epsilon-star-simplified}) is repeated for
the Green-St.~Venant strain, Eq.~(\ref{ESD-3D}), then
Eqs.~(\ref{epsilon-star-simplified}) are replaced by
$\tilde{\epsilon}_{rr}^{\star}=\tilde{\epsilon}_{rr}-2x_3
b_{rr}+x_3^2c_{rr}$, and
$\tilde{\epsilon}_{\theta\theta}^{\star}=\tilde{\epsilon}_{\theta\theta}-2x_3
b_{\theta\theta}/\Phi^2+x_3^2 c_{\theta\theta}/\Phi^2$. The different
dependence on the $x_3$ coordinate perpendicular to the mid-surface,
inevitably leads to additional terms upon integration of the energy
over $x_3$.

Quantitatively, the differences caused by the coupling term are small
and indeed may lie outside the strict limits of the infinitesimal-strain theory.
They seem negligible experimentally. The removal of this
term, however, leads to a much simpler analysis, as demonstrated by
the exact solutions in Sec.~\ref{exact-sol-planar}. (A similar
observation was made in the context of beam theory \cite{Irschik}.)
Since, at least for the problems considered in this manuscript, the differences can be neglected, there is freedom, and clear
benefit, in choosing a more tractable formulation when it is
available.

The two models become equivalent in the incompressible limit,
$B/Y=0$. The problems treated in Secs.~\ref{cylindrical-symmetry} and
\ref{self-consis} reveal an essential difference in the way the two
models {\em depart} from this limit. Both problems\,---\,the
uniaxially bent sheet and the sheet with elliptic reference
metric\,---\,possess a strain-free configuration (isometric immersion)
as the energy minimizer for $B/Y=0$. According to the ESK model this
configuration ceases to be the minimizer for an arbtirarily small but
finite $B/Y$; according to the model presented here it remains the
energy minimizer to leading order in $B/Y$. In other words, as $B/Y$
tends to zero, the equilibrium configuration reaches the isometry with
nonzero slope in the former, and with zero slope in the latter. In a
sheet made of a 3D material both $Y$ and $B$ emanate from the same
elastic modulus. Then, it may well be that a stretching-bending
coupling exists even in the absence of Gaussian curvature, leading with decreasing thickness to the ``nonzero
slope'' behavior. In a genuinely 2D sheet, such as a monomolecular
layer or a 2D polymer network, $Y$ and $B$ can be independent (e.g.,
arising from the rigidities of bonds and bond angles,
respectively). In such cases, for example, it may well be that
stretching and bending should be decoupled, leading to the ``zero
slope'' case\,---\,i.e., an isometry (no stretched bonds) remaining
the energy minimizer for $B/Y>0$ (finite joint rigidity). These
delicate issues might be checked in discrete simulations. While being
conceptually interesting, they may have (at least according to the problems considered here) little practical
significance.


The exact solutions presented in Sec.~\ref{exact-sol-planar} for the
strains and stresses in flat incompatible sheets can be used as the
base solutions for a perturbative (near-threshold) treatment of
buckling instabilities in these systems, which can then be studied
experimentally.  Our formulation can be applied to additional examples
beyond those addressed in Secs.~\ref{exact-sol-planar}
and~\ref{self-consis}, where the reference metric is axisymmetric. An
interesting problem, for instance, might be the case of a highly
localized (delta-function) $\Phi(r)$.  In addition, the theory might
be useful for analyzing stress fields around two-dimensional defects
\cite{moshe2,moshe1}. 

The most important extension of this work, however, would be to obtain
a similarly tractable formulation for sheets of any two-dimensional
deformation. The discussion above suggests two possible routes. One is
to generalize the formulation presented in Sec.~\ref{2D-alt-model}
beyond axisymmetric deformations. The other is to modify the ESK
energy functional such that the two choices of strain measures lead to
equivalent theories.

\begin{acknowledgments}
We are indebted to Efi Efrati, Eran Sharon, and Raz Kupferman for
extensive, illuminating discussions. We thank James Hanna, Robert Kohn,
Michael Moshe, and Tom Witten for helpful comments. This work has been
supported in part by the Israel Science Foundation (Grant No. 164/14).
\end{acknowledgments}

\appendix

\section{Consistent energy minimization for a uniaxially deformed sheet}
\label{app-1D-minimization}

In this Appendix we show that minimization of $E_{\rm 1D}$ with
respect to $\epsilon_{ss}$ and $\phi$ yields equations of equilibrium
which are identical to the ones obtained by the appropriate
minimization with respect to the spatial configuration, ${\bf f}(s)$.

We first define the perturbed configuration, $\tilde{\bf f}(s)$, by
\begin{equation}
\tilde{\bf f}(s)={\bf f}(s)+\delta{\bf f}(s)={\bf f}(s)+\psi_t(s){\bf \hat{t}}+\psi_n(s){\bf \hat{n}},
\label{configuration}
\end{equation}
where $\{{\bf \hat{t}}(s),{\bf \hat{n}}(s)\}$ are the unit vectors
tangent and normal to the sheet along the deformation axis, and
$\psi_t$ and $\psi_n$ are arbitrary perturbation
functions. Equivalently (up to a shift of the origin), we can represent
the configuration by $d{\bf f}/ds$, i.e., $d\tilde{\bf f}/ds = d{\bf
  f}/ds + d\delta{\bf f}/ds$. Then, the variation of the energy is
written as
\begin{equation}
  \delta E_{\rm 1D}=\int({\cal E}_t{\bf \hat{t}}+{\cal E}_n{\bf
    \hat{n}})\cdot\frac{d\delta{\bf f}}{ds}ds,
\label{deltaE1D}
\end{equation}
where ${\cal E}_t$ and ${\cal E}_n$ are some functions of
$\epsilon_{ss}$ and $\phi$ yet to be determined. We wish to relate
the variation $d\delta{\bf f}/ds$ with the variations
$\delta\epsilon_{ss}$ and $\delta\phi$.

The vectors $\{{\bf \hat{t}}(s),{\bf \hat{n}}(s)\}$ satisfy the
Frenet-Serret formulas \cite{lipschutz},
\begin{subequations}\label{F-S}
\begin{eqnarray}
&&\frac{d{\bf \hat{t}}}{ds}=(1+\epsilon_{ss})\kappa{\bf \hat{n}}=\frac{d\phi}{ds}{\bf \hat{n}}, \label{F-S-1} \\
&&\frac{d{\bf \hat{n}}}{ds}=-(1+\epsilon_{ss})\kappa{\bf \hat{t}}=-\frac{d\phi}{ds}{\bf \hat{t}}, \label{F-S-2}
\end{eqnarray}
\end{subequations}
where $\kappa=d\phi/d\hat{s}$ is the curvature and $\hat{s}$ is the
arclength of the deformed configuration,
$d\hat{s}/ds=1+\epsilon_{ss}$. With the help of Eqs.~(\ref{F-S}),
differentiating $\delta{\bf f}$ of Eq.~(\ref{configuration}) with
respect to $s$ gives
\begin{equation}\label{deltaRderi}
\frac{d\delta{\bf
    f}}{ds}=\left(\frac{d\psi_t}{ds}-\frac{d\phi}{ds}\psi_n\right){\bf
  \hat{t}}+\left(\frac{d\psi_n}{ds}+\frac{d\phi}{ds}\psi_t\right){\bf
  \hat{n}}.
\end{equation} 
Next, we examine the in-plane variation $\delta\epsilon_{ss}$ to
leading order in the perturbation functions, 
\begin{equation}\label{delta-epsiln-ss-fi}
\delta\epsilon_{ss}=
\left|\frac{d{\bf \tilde{f}}}{ds}\right|-\left|\frac{d{\bf f}}{ds}\right|\simeq \frac{d\psi_t}{ds}-\frac{d\phi}{ds}\psi_n,
\end{equation}
To do the same for the $\delta\phi$ we start by writing $\cos\phi={\bf
  \hat{t}}\cdot{\bf \hat{x}}$, where ${\bf \hat{x}}$ is a constant
unit vector along the horizontal direction. Upon variation we have,
$-\sin\phi\delta\phi=\delta{\bf \hat{t}}\cdot{\bf \hat{x}}$. In turn, the
variation of the tangent vector is given by,
\begin{equation}
\delta{\bf \hat{t}}=\frac{d{\bf \tilde{f}}/ds}{|d{\bf
    \tilde{f}}/ds|}-\frac{d{\bf f}/ds}{|d{\bf
    f}/ds|}\simeq\frac{1}{1+\epsilon_{ss}}\left(\frac{d\psi_n}{ds}+\frac{d\phi}{ds}\psi_t\right){\bf
  \hat{n}},
\end{equation}
and, since ${\bf \hat{n}}\cdot{\bf \hat{x}}=-\sin\phi$, we get
\begin{equation}\label{delta-phi-fi}
(1+\epsilon_{ss})\delta\phi=\frac{d\psi_n}{ds}+\frac{d\phi}{ds}\psi_t.
\end{equation}
Collecting the results for $\delta\epsilon_{ss}$ and $\delta\phi$
(Eqs.~(\ref{delta-epsiln-ss-fi}) and (\ref{delta-phi-fi})) and
substituting in Eq.~(\ref{deltaRderi}), we obtain the desired relation,
\begin{equation}
  \frac{d\delta{\bf f}}{ds}= \delta\epsilon_{ss} {\bf
    \hat{t}}+(1+\epsilon_{ss})\delta\phi{\bf \hat{n}}.
\end{equation}
This proves that the variation with respect to the spatial
configuration is equivalent to the variation with respect to
$\delta\epsilon_{ss}$ and $\delta\phi$.

We can proceed to rewrite the variation of the energy,
Eq.~(\ref{deltaE1D}), as 
\begin{equation}\label{var-energy-global}
\delta E_{\rm 1D}=\int\left[{\cal
    E}_t\delta\epsilon_{ss}+(1+\epsilon_{ss}){\cal
    E}_n\delta\phi\right]ds.
\end{equation}
The straightforward way to get the equations of equilibrium is to set
this functional to zero for arbitrary $\delta\epsilon_{ss}$ and
$\delta\phi$, i.e., ${\cal E}_t=0$ and ${\cal E}_n=0$. This is what
has been done in Sec.~\ref{cylindrical-symmetry}, where
\begin{subequations}\label{global-eqn-ex}
\begin{eqnarray}
&&{\cal E}_t=Y\epsilon_{ss}=\sigma_{ss} = 0,\label{global-eqn-ex-1} \\
&&{\cal E}_n=-\frac{B}{1+\epsilon_{ss}}\frac{d^2\phi}{ds^2}=-\frac{dM_{ss}}{d\hat{s}} = 0.
\end{eqnarray}
\end{subequations}
(See Eqs.~(\ref{sigmaM}) and (\ref{1D-eqn-M-sigma}).)

Alternatively, we can rewrite the energy variation,
Eq.~(\ref{deltaE1D}), in terms of $\delta{\bf f}$ rather than
$d\delta{\bf f}/ds$, using intergration by parts. This yields the
equations of equilibrium in the different form,
\begin{subequations}\label{loacl-eqn-diff}
\begin{eqnarray}
&&\frac{d{\cal E}_t}{ds}-\frac{d\phi}{ds}{\cal E}_n=0, \\
&&\frac{d{\cal E}_n}{ds}+\frac{d\phi}{ds}{\cal E}_t=0.
\end{eqnarray}
\end{subequations} 
Subtituting Eqs.~(\ref{global-eqn-ex}), this gives
\begin{subequations}\label{local-eqn-ex}
\begin{eqnarray}
&&\frac{d\sigma_{ss}}{d\hat{s}}-\kappa\sigma_{sn}=0, \\
&&\frac{d\sigma_{sn}}{d\hat{s}}+\kappa\sigma_{ss}=0,
\end{eqnarray}
\end{subequations}
where $\sigma_{sn}=-dM_{ss}/d\hat{s}$ is the normal force at a cross
section \cite[p.~387]{Love}.

\begin{figure}[!tbh]
\vspace{0.7cm}
  \centering
  \includegraphics[width=13cm]{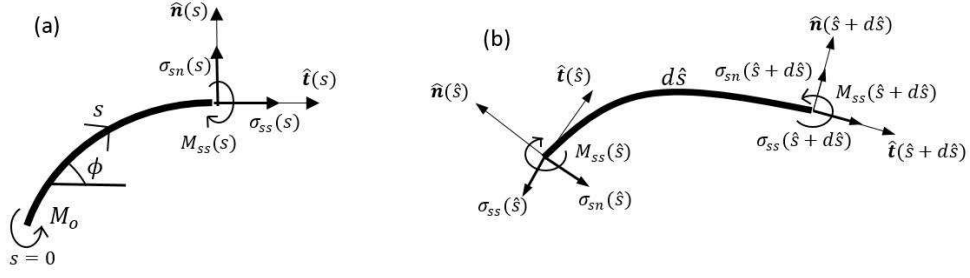} \ \ \ 
	\caption{(a) Schematic force balance on a finite sheet
          segment. A bending moment, $M_o$, applied at the
          boundary, is balanced by the reaction forces, $\sigma_{ss}$
          and $\sigma_{sn}$, and the reaction bending moment,
          $M_{ss}$. Under these conditions $\sigma_{ss}=\sigma_{sn}=0$
          and $M_{ss}=M_o$, consistently with
          Eqs.~(\ref{global-eqn-ex}).  (b) Schematic force balance on
          an infinitesimal sheet segment of length
          $d\hat{s}$. Balance of forces in the tangential direction,
          ${\bf\hat{t}}(\hat{s})$, is given by,
          $-\sigma_{ss}(\hat{s})+\sigma_{ss}(\hat{s}+d\hat{s}){\bf
            \hat{t}}(\hat{s}+d\hat{s})\cdot{\bf
            \hat{t}}(\hat{s})+\sigma_{sn}(\hat{s}+d\hat{s}){\bf
            \hat{n}}(\hat{s}+d\hat{s})\cdot{\bf
            \hat{t}}(\hat{s})=0$. Expanding this equation to leading
          order in the differential $d\hat{s}$ (using
          Eqs.~(\ref{F-S})) we obtain
          $d\sigma_{ss}/d\hat{s}-\kappa\sigma_{sn}=0$. Similarly,
          force balance in the normal direction and balance of bending
          moments gives: $d\sigma_{sn}/d\hat{s}+\kappa\sigma_{ss}=0$
          and $dM_{ss}/d\hat{s}+\sigma_{ns}=0$, consistently with
          Eqs.~(\ref{local-eqn-ex}).}
  \label{global-local-force}
\end{figure}

The difference between the two equivalent sets of equilibrium
equations is explained in Fig.~\ref{global-local-force}. While
Eqs.~(\ref{global-eqn-ex}) represent balance of forces and moments
across a finite segment of the sheet, Eqs.~(\ref{local-eqn-ex})
represent the balance for an infinitesimal segment.

\section{Boundary layer in a sheet with elliptic reference metric}
\label{boundary-layer-elliptic}

In this Appendix we show that the energy of the isometric spherical
cap, Eq.~(\ref{f-shape-elliptic}), is reduced when a boundary layer is
formed (i) near the outer radius of a complete disc, and (ii) near the outer and inner radii of an annulus. 
The existence of these boundary layers and
the scaling of their width with the thickness $t$ were found in
Ref.~\cite{Efi2}. Here we derive these results based on a variational
Ansatz within the FvK approximation, thus obtaining full expressions
including prefactors.

Considering the elliptic reference metric, Eq.~(\ref{Phi-elliptic}),
and employing the small-slope approximation, we obtain for the
in-plane strains, Eqs.~(\ref{epsilon-in-plane}),
\begin{subequations}\label{strain-FvK-bl}
\begin{eqnarray}
&&\epsilon_{rr}\simeq\partial_r u_r+\frac{1}{2}(\partial_r\zeta)^2, \label{epsilon-rr-FvK} \\
&&\epsilon_{\theta\theta}\simeq \frac{Kr^2}{6}+\frac{u_r}{r}.  \label{epsilon-2theta-FvK}
\end{eqnarray}
\end{subequations}
For the isometric immersion these strains vanish, yielding the height function $\zeta_{iso}\simeq\sqrt{K}r^2/2$. The total energy of the spherical cap is obtained by substituting this function in Eq.~(\ref{Eb-energy-f-FvK1}), giving, 
\begin{equation}
E_{iso}=\pi (1+\nu)(KR^2)B.
\label{E-iso-FvK}
\end{equation}

Let us try to reduce the total energy below $E_{iso}$ through the following variational Ansatz:
\begin{equation}
\zeta(r)=\zeta_{iso}+\zeta_{bl}=\frac{\sqrt{K}r^2}{2}-\frac{(1+\nu)}{\alpha(\alpha+\nu-1)}\sqrt{K}R^2\left(\frac{r}{R}\right)^{\alpha},
\label{height-bl}
\end{equation}
where $\alpha$ serves as a variational parameter. The coefficient of the second term in Eq.~(\ref{height-bl}) has been chosen so as to satisfy the boundary condition of zero radial bending moment at the outer radius, $\left.M_{rr}\right|_{r=R}\simeq B\left[\partial_{rr}\zeta+\frac{\nu}{r}\partial_r\zeta\right]_{r=R}=0$. When $\alpha\gg 1$ the additional term is negligible everywhere except close to the edge, as expected from a boundary layer. As shown below, the minimizing configuration has $\alpha\sim t^{-1/2}$. 

Since our Ansatz, Eq.~(\ref{height-bl}), is not an isometry, it contains in-plane stress. To calculate this stress we first minimize the stretching energy, Eq.~(\ref{Es-final-sigma}), with respect to $u_r$. In the FvK approximation the resulting equation reads,
\begin{equation}
\partial_r(r\sigma_{rr})-\sigma_{\theta\theta}=0.
\label{in-plane-stress-FvK}
\end{equation}
Substituting, Eq.~(\ref{height-bl}) in the strains, Eqs.~(\ref{strain-FvK-bl}), and then in the stress-strain relations, Eqs.~(\ref{sigma-2D}), we obtain from Eq.~(\ref{in-plane-stress-FvK}),
\begin{equation}
r\partial_r(r\partial_r u_r)-u_r=-\frac{4}{3}Kr^3+(1+\nu)\frac{\alpha-\nu+1}{\alpha+\nu-1}KR^3\left(\frac{r}{R}\right)^{\alpha+1}+\frac{1}{2}(1+\nu)^2\frac{\nu-2\alpha+1}{(\alpha+\nu-1)^2}KR^3\left(\frac{r}{R}\right)^{2\alpha-1}.
\label{ur-eqn-FvK-bl}
\end{equation}
Two boundary conditions are necessary: one is a vanishing stress at the free edge, $\left.\sigma_{rr}\right|_{r=R}=0$, and the other  is a vanishing displacement at the origin, $\left.u_r\right|_{r=0}=0$. 
The solution of Eq.~(\ref{ur-eqn-FvK-bl}) subject to these conditions is,
\begin{equation}
u_r(r)=A_0 r-\frac{K}{6}r^3+\frac{(1+\nu)(\alpha-\nu+1)}{\alpha(\alpha+2)(\alpha+\nu-1)}KR^3\left(\frac{r}{R}\right)^{\alpha+1}+\frac{1}{8}\frac{(1+\nu)^2(\nu-2\alpha+1)}{\alpha(\alpha-1)(\alpha+\nu-1)^2}KR^3\left(\frac{r}{R}\right)^{2\alpha-1},
\label{ur-sol-FvK-iso}
\end{equation}
where $A_0$ is determined by the first boundary condition.

Substituting $u_r$ and $\zeta$ from  Eqs.~(\ref{height-bl}) and (\ref{ur-sol-FvK-iso}), in Eqs.~(\ref{Es-final-sigma}) and (\ref{Eb-energy-f-FvK1}), and expanding to leading order in $1/\alpha$, gives,
\begin{equation}
E\simeq \frac{\pi}{2}YR^2(KR^2)^2(1-\nu)(1+\nu)^3\alpha^{-5}+\pi(1+\nu)(KR^2)B-\frac{3\pi}{2}(1+\nu)^2(KR^2)B\alpha^{-1},
\label{E-approx-alpha}
\end{equation}
where the first term comes from stretching and the last two are bending contributions. Minimization of Eq.~(\ref{E-approx-alpha}) with respect to $\alpha$ yields,
\begin{equation}
\alpha=(5/3)^{1/4}(1-\nu^2)^{1/4}(KR^2)^{1/4}(YR^2/B)^{1/4}=(20)^{1/4}(1-\nu^2)^{1/4}(KR^2)^{1/4}(t/R)^{-1/2}.
\label{alpha-sol-iso}
\end{equation}
Substituting this result in Eq.~(\ref{E-approx-alpha}) we finally obtain,
\begin{equation}
E\simeq E_{iso}-\frac{6\pi}{5}\left(\frac{3}{5}\right)^{1/4}\frac{(1+\nu)^2}{(1-\nu^2)^{1/4}}(KR^2)^{3/4}\left(\frac{B}{YR^2}\right)^{1/4}B,
\end{equation}
where $E_{iso}$ is given by Eq.~(\ref{E-iso-FvK}). Thus, the energy of the isometric immersion is reduced by the introduction of a boundary layer. The reduction scales as $t^{7/2}$ whereas $E_{iso}\sim t^3$.  In the limit of small thickness we can write  $\zeta_{bl}(r)\simeq -\frac{(1+\nu)\sqrt{K}R^2}{\alpha^2}e^{-(R-r)/w}$ with the width of the boundary layer being,
\begin{equation}
w=R/\alpha=(20)^{-1/4}(1-\nu^2)^{-1/4}(KR^2)^{-1/4}(t/R)^{1/2}R.
\end{equation}

This derivation can straightforwardly be extended to the more general case of an annulus with inner radius $R_i$ and outer radius $R_o$. In this case the energy of the isometric immersion, $\zeta_{iso}$, is given by,
\begin{equation}\label{E-iso-annulus-bl}
E_{iso}=\pi(1+\nu)K(R_o^2-R_i^2)B.
\end{equation} 
This energy can be reduced below $E_{iso}$ if two boundary layers are formed near the outer and inner radii, as indicated by the following Ansatz,
\begin{equation}
\zeta(r)=\zeta_{iso}+\zeta_{bl}=\frac{\sqrt{K}r^2}{2}+A_o\left(\frac{r}{R_o}\right)^{\alpha}+B_o\left(\frac{R_i}{r}\right)^{\alpha}.
\end{equation}
As in the case of a disc, $A_o$ and $B_o$ are chosen such that the radial bending moment is zero at the two boundaries, $\left.M_{rr}\right|_{r=R_i,R_o}=0$. This gives,
\begin{subequations}
\begin{eqnarray}
&&A_o=-\sqrt{K}R_o^2\frac{1+\nu}{\alpha(\alpha+\nu-1)}\frac{1-\rho^{\alpha+2}}{1-\rho^{2\alpha}}, \\
&&B_o=-\sqrt{K}R_i^2\frac{1+\nu}{\alpha(\alpha-\nu+1)}\frac{1-\rho^{\alpha-2}}{1-\rho^{2\alpha}},
\end{eqnarray}
\end{subequations} 
where $\rho\equiv R_i/R_o$.

Following the same route as in Eqs.~(\ref{in-plane-stress-FvK})-(\ref{ur-sol-FvK-iso}), we find after expansion in powers of $\alpha^{-1}$ and assuming  $\rho^{\alpha}\rightarrow 0$ that the total energy of the annulus is given by, 
\begin{eqnarray}\label{energy-bl-annulus}
E\simeq \frac{\pi}{2}YR_o^2(K R_o^2)^2\left(1+\rho^6\right)(1-\nu)(1+\nu)^3\alpha^{-5}+\pi(1+\nu)K(R_o^2-R_i^2)B 
-\frac{3\pi}{2}(1+\nu)^2(K R_o^2)\left(1+\rho^2\right)B\alpha^{-1}. \nonumber \\
\end{eqnarray}
Minimization of this energy with respect to $\alpha$ gives,
\begin{eqnarray}\label{C15}
\alpha&=&(5/3)^{1/4}(1-\nu^2)^{1/4}\left(\frac{1+\rho^6}{1+\rho^2}\right)^{1/4}(K R_o^2)^{1/4}\left(\frac{Y R_o^2}{B}\right)^{1/4}\nonumber \\ &=&(20)^{1/4}(1-\nu^2)^{1/4}\left(\frac{1+\rho^6}{1+\rho^2}\right)^{1/4}(K R_o^2)^{1/4}\left(\frac{t}{R_o}\right)^{-1/2}.
\end{eqnarray}
Note that in the limit of $\rho\rightarrow 0$ this result coincides with Eq.~(\ref{alpha-sol-iso}). Substituting Eq.~(\ref{C15}) back in the energy, Eq.~(\ref{energy-bl-annulus}), we obtain,
\begin{equation}
E\simeq E_{iso}-\frac{6\pi}{5}\left(\frac{3}{5}\right)^{1/4}\frac{(1+\nu)^2}{(1-\nu^2)^{1/4}}(KR_o^2)^{3/4}\frac{\left(1+\rho^2\right)^{5/4}}{\left(1+\rho^6\right)^{1/4}}\left(\frac{B}{YR_o^2}\right)^{1/4}B,
\end{equation}
where $E_{iso}$ is given by Eq.~(\ref{E-iso-annulus-bl}). Thus, the introduction of two boundary layers, at the inner and outer radii of the annulus, reduce the energy of an isometric immersion.

\section{Stability criterion for isometric immersions with negative Gaussian curvature}
\label{negative-Gauss-curv}
In this appendix we extend the theory presented in Sec.~\ref{2D-alt-model} to surfaces of revolution, [see Eq.~(\ref{f-shape})], whose reference metric is given by,
\begin{equation}
\bar{g}_{\alpha\beta}=
\begin{pmatrix}
\bar{g}_{r}^2 & 0 \\ 0 & \bar{g}_{\theta}^2
\end{pmatrix}
\ \ \ , \ \ \ ds^2=\bar{g}_{r}^2(r)dr^2+\bar{g}_{\theta}^2(r)d\theta^2.
\label{ref-metric2}
\end{equation}
Our aim is to derive a self-consistent stability criterion, similar to Eqs.~(\ref{sel-cosis-eom-bc}), for isometric immersions with constant negative Gaussian curvature \cite{gemmer}. 

Following Sec.~\ref{2D-alt-model} it is straightforward to show that the energy functional, Eq.~(\ref{E-2D-final}), is modified into,
\begin{equation}
E=\frac{Y}{2}\int_0^{R}\int_0^{2\pi}\left[\epsilon_{rr}^2+\epsilon_{\theta\theta}^2+2\nu\epsilon_{rr}\epsilon_{\theta\theta}\right]\bar{g}_{r}\bar{g}_{\theta} d\theta dr+\frac{B}{2}\int_0^{R}\int_0^{2\pi}\left[\phi_{rr}^2+\phi_{\theta\theta}^2+2\nu\phi_{rr}\phi_{\theta\theta}\right]\bar{g}_{r}\bar{g}_{\theta} d\theta dr,
\label{E-2D-final-B}
\end{equation}
where the in-plane strains are given by,
\begin{subequations}\label{strains-2D-B}
\begin{eqnarray}
&&\epsilon_{rr}=\sqrt{a_{rr}}/\bar{g}_{r}-1=\sqrt{(1+\partial_r u_r)^2+(\partial_r\zeta)^2}/\bar{g}_{r}-1, \label{epsilon-rr-B} \\
&&\epsilon_{\theta\theta}=\sqrt{a_{\theta\theta}}/\bar{g}_{\theta}-1=(r+u_r)/\bar{g}_{\theta}-1, \label{epsilon-2theta-B}
\end{eqnarray}
\end{subequations}
and the ``bending-strains", are given by,
\begin{subequations}\label{bending-strains-2D-B}
\begin{eqnarray}
&&\phi_{rr}=\sqrt{c_{rr}}/\bar{g}_{r}=\frac{1}{\bar{g}_{r}}\frac{(1+\partial_r u_r)\partial_{rr}\zeta-\partial_{rr}u_r\partial_r\zeta}{(1+\partial_r u_r)^2+(\partial_r\zeta)^2}=\partial_r\phi^r/\bar{g}_{r}, \label{phi-rr-B} \\
&&\phi_{\theta\theta}=\sqrt{c_{\theta\theta}}/\bar{g}_{\theta}=\frac{1}{\bar{g}_{\theta}}\frac{\partial_r\zeta}{\sqrt{(1+\partial_r u_r)^2+(\partial_r\zeta)^2}}=\sin\phi^r/\bar{g}_{\theta}. \label{phi-2theta-B}
\end{eqnarray}
\end{subequations}

Setting Eqs.~(\ref{strains-2D-B}) to zero, we obtain the displacement corresponding to the isometric immersion of Eq.~(\ref{ref-metric2}),
\begin{subequations}\label{iso-immer-B}
\begin{eqnarray}
&&u_r=\bar{g}_{\theta}-r, \label{ur-iso-B} \\
&&\partial_r\zeta=\sqrt{\bar{g}_{r}^2-(\partial_r \bar{g}_{\theta})^2} \label{zeta-is-B}.
\end{eqnarray}
\end{subequations}
Following the analysis in Sec.~\ref{self-consis}, we minimize the bending energy,
\begin{equation}
E_b=\frac{1}{2}\int_0^R\int_0^{2\pi}\left[M_{rr}\partial_r\phi^r/\bar{g}_{r}+M_{\theta\theta}\sin\phi^r/\bar{g}_{\theta}\right]\bar{g}_{r}\bar{g}_{\theta}d\theta dr, \nonumber
\end{equation}
with respect to $\phi^r$ to obtain the balance of bending moments. This gives,
\begin{eqnarray}
\partial_r(\bar{g}_{\theta}M_{rr})-\bar{g}_{r}\cos\phi^r M_{\theta\theta}=0. \label{moment-balance-B} 
\end{eqnarray}
where $M_{\alpha\beta}$ are given by Eqs.~(\ref{M-2D}) and  $\phi_{\alpha\beta}$ are given by Eqs.~(\ref{bending-strains-2D-B}).

Substituting the displacements of Eqs.~(\ref{iso-immer-B}) in the ``bending strains", Eqs.~(\ref{bending-strains-2D-B}), we obtain,
\begin{subequations}\label{phi-2D-B-iso}
\begin{eqnarray}
&&\phi_{rr}=\left(\partial_r\bar{g}_{\theta}\partial_r\sqrt{\bar{g}_r^2-(\partial_r\bar{g}_{\theta})^2}-\partial_{rr}\bar{g}_{\theta}\sqrt{\bar{g}_r^2-(\partial_r\bar{g}_{\theta})^2}\right)/\bar{g}_r^3, \label{phi-rr-iso-B} \\
&&\phi_{\theta\theta}=\sqrt{\bar{g}_r^2-(\partial_r\bar{g}_{\theta})^2}/(\bar{g}_r\bar{g}_{\theta}). \label{phi-2theta-iso-B}
\end{eqnarray}
\end{subequations}
In addition, using Eq.~(\ref{phi-2theta-B}), we have that,
\begin{equation}
\cos\phi^r=\partial_r\bar{g}_{\theta}/\bar{g}_r. \label{cos-phi-B}
\end{equation}
Substituting Eqs.~(\ref{phi-2D-B-iso}) in (\ref{M-2D}) and then, along with Eq.~(\ref{cos-phi-B}), in (\ref{moment-balance-B}) we finally obtain the self-consistency condition,
\begin{eqnarray}
\partial_r\left(\bar{g}_{\theta}\left(\partial_r\bar{g}_{\theta}\partial_r\sqrt{\bar{g}_r^2-(\partial_r\bar{g}_{\theta})^2}-\partial_{rr}\bar{g}_{\theta}\sqrt{\bar{g}_r^2-(\partial_r\bar{g}_{\theta})^2}\right)/\bar{g}_r^3\right)-\partial_r\bar{g}_{\theta}\sqrt{\bar{g}_r^2-(\partial_r\bar{g}_{\theta})^2}/(\bar{g}_r\bar{g}_{\theta})=0.
\label{self-consis-B}
\end{eqnarray}

It is now straightforward to verify that a pseudosphere,
$\bar{g}_r=\tanh r$ and $\bar{g}_{\theta}=1/\cosh r$, and hyperboloid
of revolution, $\bar{g}_r=b\ \text{sn}(r,b)$ and
$\bar{g}_{\theta}=\text{dn}(r,b)$ ($\text{sn}$ and $\text{dn}$
denoting the Jacobi elliptic functions \cite{Abramowitz}), both do not
satisfy Eq.~(\ref{self-consis-B}). Thus, both are mechanically
unstable. As in the case of the cone, we note that these conclusions
do not rule out the possibility that the objects approach these shapes
in the limit $t\rightarrow 0$.

\section{Supplementary material:\\  Comparison between thin sheet theories based on model-independent force-balance equations}
\label{Supplementary}

As has been demonstrated in the main text by several examples, the ESK
model and the present one produce different equations of equilibrium
and 
different equilibrium configurations. Yet, obviously,
both models describe balance of forces and torques. Therefore, using
an appropriate representation, both should result in identical (albeit
not equivalent) equations of equilibrium. Thus the lack of equivalence
would be confined to the relations between stress and deformation (the
constitutive relations), and we would get an instructive comparison of
the stress and torque under similar loading conditions in the two
models. Such a representation is the goal of this Supplemental
Material.

While the present model is based on the Biot strain measure,
Eq.~(\ref{epsilon-star}), the ESK theory \cite{Efi1} is based on the
second Piola-Kirchhoff strain, Eq.~(\ref{ESD-3D}). As a result, our
equilibrium equations, Eqs.~(\ref{eom-flat-ur}) and
(\ref{moment-balance-eom}), manifestly differ from the ones obtained
in Ref.~\cite{Efi1}, Eq.~(3.10) in that paper.
  
To derive the conditions of force and torque balance we first define
the co-moving coordinate system $\{{\bf \hat{t}}_r,{\bf
  \hat{t}}_{\theta},{\bf \hat{n}}\}$, where ${\bf
  \hat{t}}_{\alpha}=\partial_{\alpha}{\bf f}/|\partial_{\alpha}{\bf
  f}|$ are two in-plane unit vectors and ${\bf \hat{n}}$ is the unit
normal, given the spatial configuration ${\bf f}(r,\theta)$.  Second,
we cut an infinitesimal patch of the surface, whose borders lie along
lines of constant coordinates \cite[p.~24]{Goldenveizer}, and balance
the force and torque vectors applied on its edges. This gives
\cite[p.~29]{Goldenveizer},
\begin{subequations}\label{force-torque-balance-2D}
\begin{eqnarray}
&&0=\partial_r(\Phi {\bf F}_r)+\partial_{\theta}{\bf F}_{\theta},\label{force-torque-balance-2D-1} \\
&&0=\partial_r(\Phi {\bf M}_r)+\partial_{\theta}{\bf M}_{\theta}-\Phi\partial_r{\bf f}\times{\bf F}_r-\partial_{\theta}{\bf f}\times{\bf F}_{\theta}, \label{force-torque-balance-2D-2}
\end{eqnarray}
\end{subequations}
where ${\bf F}_{\alpha}$ and ${\bf M}_{\alpha}$ are the forces and
bending moments per undeformed unit length along the directions
$\alpha=r,\theta$. Lastly, we resolve the components of these vectors
projected on our triad basis,
\begin{subequations}\label{stress-torque-com}
\begin{eqnarray}
&&{\bf F}_{\alpha}=\sigma_{\alpha r}{\bf \hat{t}}_r+\sigma_{\alpha\theta}{\bf \hat{t}}_{\theta}+\sigma_{\alpha 3}{\bf \hat{n}}, \label{stress-torque-com-1}\\
&&{\bf M}_{\alpha}={\bf \hat{n}}\times(M_{\alpha r}{\bf \hat{t}}_r+M_{\alpha \theta}{\bf \hat{t}}_{\theta}). \label{stress-torque-com-2}
\end{eqnarray}
\end{subequations}

Note the delicate point, crucial for the sake of this section, that
the tensors $\sigma_{\alpha\beta}$ and $M_{\alpha\beta}$ here
correspond to the {\em actual} forces and torques, i.e., the fluxes of
linear and angular momenta. As such, they do not depend on the choice
of model; unlike Eqs.~(\ref{sigma-2D}) and (\ref{M-2D}) in the main
text, we do not relate them at this moment to a certain definition of
strain. In other words, they are not necessarily equal to the
variation of the energy of the chosen model with respect to the strain
and curvature of that model. Similarly, the configuration is
represented in these equations through the model-independent spatial
triad $\{{\bf \hat{t}}_r,{\bf \hat{t}}_{\theta},{\bf \hat{n}}\}$.

For axisymmetric deformations, Eq.~(\ref{f-shape}), we always have
$\sigma_{r\theta}=M_{r\theta}=0$, and
Eqs.~(\ref{force-torque-balance-2D}) and (\ref{stress-torque-com})
form a system of five differential equations for the eight unknowns,
$\{\sigma_{\alpha\alpha},\sigma_{\alpha
  3},M_{\alpha\alpha},u_r,\zeta\}$, where the configuration is now
represented by the displacements $u_r$ and $\zeta$, obtainable from
$\{{\bf \hat{t}}_r,{\bf \hat{t}}_{\theta},{\bf \hat{n}}\}$. Thus, to
have a closure we must derive constitutive relations between the
stress and torque components and the actual deformation. 

The definition of mechanical energy, as well, does not depend on the
choice of model. It is the sum of two terms: (i) The work  done by in-plane forces to  displace the sheet from its rest state to the given configuration (not displacement squared), and (ii) the work done by bending moments to change the out-of-plane angles from their rest values.  For clarity of the expressions that follow, it is
helpful to represent the displacements equivalently by in-plane
stretching fields,
$\gamma_{\alpha\alpha}\equiv\sqrt{a_{\alpha\alpha}/\bar{g}_{\alpha\alpha}}$,
and out-of-plane bending fields, $\bar{g}_{\alpha\alpha}\phi_{\alpha\alpha}\equiv
b_{\alpha\alpha}/\gamma_{\alpha\alpha}$ (where $\alpha=r,\theta$, and
the mixed terms vanish by axisymmetry). The variation of the energy is
given then by the infinitesimal work,
\begin{equation}\label{delta-E-model-free}
  \delta E = \int_{0}^R\int_{0}^{2\pi}
  \left(\sigma_{\alpha\beta}\delta\gamma_{\alpha\beta} + M_{\alpha\beta}\delta\phi_{\alpha\beta} \right)
  \Phi dr d\theta.
\end{equation}
We note that Eq.~(\ref{delta-E-model-free}) is the 2D extension of the so-called principle of virtual work \cite{Irschik,Reissner1972}. In addition, similar to our proof in Appendix~A it can be shown that $\delta\gamma_{\alpha\beta}$ and $\delta\phi_{\alpha\beta}$ are consistent with minimization of the energy with respect to the configuration. These infinitesimals are proportional to the 1D variations, $\delta\epsilon_{ss}$ and $\delta\phi$, considered in Sec.~\ref{cylindrical-symmetry}.

If we now consider the energy functional of each model, express it in
terms of the actual deformation fields $\gamma_{\alpha\alpha}$ and
$\phi_{\alpha\alpha}$, and take the variation with respect to these
fields, we will get the constitutive relations for the actual
stresses and bending moments, as arising from each model.

The energy functional of the present model
(Eq.~(\ref{E-2D-final})) is rewritten in terms of the
deformation fields as
\begin{eqnarray}
E_{\rm 2D} = &&
\frac{Y}{2}\int_0^{R}\int_0^{2\pi}\left[(\gamma_{rr}-1)^2+(\gamma_{\theta\theta}-1)^2
  +2\nu(\gamma_{rr}-1)(\gamma_{\theta\theta}-1)\right]\Phi d\theta dr \nonumber\\
  &&+\frac{B}{2}\int_0^{R}\int_0^{2\pi}\left[\phi_{rr}^2+\phi_{\theta\theta}^2+2\nu\phi_{rr}\phi_{\theta\theta}\right]\Phi
d\theta dr.
\label{E-2D-deform}
\end{eqnarray}
Variations with respect to $\gamma_{\alpha\alpha}$ and $\phi_{\alpha\alpha}$ give
\begin{subequations}\label{cons-rel-Bio}
\begin{eqnarray}
&&\sigma_{rr}=Y\left[(\gamma_{rr}-1)+\nu(\gamma_{\theta\theta}-1)\right], \label{cons-rel-Bio-1} \\
&&\sigma_{\theta\theta}=Y\left[(\gamma_{\theta\theta}-1)+\nu(\gamma_{rr}-1)\right], \label{cons-rel-Bio-2} \\
&&M_{rr}=B\left(\phi_{rr}+\nu\phi_{\theta\theta}\right), \label{cons-rel-Bio-3} \\
&&M_{\theta\theta}=B\left(\phi_{\theta\theta}+\nu\phi_{rr}\right). \label{cons-rel-Bio-4}
\end{eqnarray}
\end{subequations}

The energy functional of the ESK model is obtained by specializing
Eq.~(\ref{ESK-2D}) to the axisymmetric case and re-expressing it
in terms of the deformation fields, yielding
\begin{eqnarray}\label{energy-ESK-uniaxial}
\mbox{ESK:}\ \ \ E_{\rm 2D}&=&\frac{Y}{8}\int_0^R\int_0^{2\pi}\left[(\gamma_{rr}^2-1)^2+(\gamma_{\theta\theta}^2-1)^2+2\nu(\gamma_{rr}^2-1)(\gamma_{\theta\theta}^2-1)\right]\Phi d\theta dr \nonumber \\
&&+\frac{B}{2}\int_0^R\int_0^{2\pi}\left[(\gamma_{rr}\phi_{rr})^2+(\gamma_{\theta\theta}\phi_{\theta\theta})^2+2\nu(\gamma_{rr}\phi_{rr})(\gamma_{\theta\theta}\phi_{\theta\theta})\right]\Phi d\theta dr.
\end{eqnarray}
Variations of this energy with respect to $\gamma_{\alpha\alpha}$ and
$\phi_{\alpha\alpha}$ give
\begin{subequations}\label{cons-rel-ESK}
\begin{eqnarray}
\mbox{ESK:}\ \ \ 
&&\sigma_{rr}=\frac{Y}{2}\gamma_{rr}\left[(\gamma_{rr}^2-1)+\nu(\gamma_{\theta\theta}^2-1)\right]+B\phi_{rr}\left(\gamma_{rr}\phi_{rr}+\nu\gamma_{\theta\theta}\phi_{\theta\theta}\right), \label{cons-rel-ESK-1} \\
&&\sigma_{\theta\theta}=\frac{Y}{2}\gamma_{\theta\theta}\left[(\gamma_{\theta\theta}^2-1)+\nu(\gamma_{rr}^2-1)\right]+B\phi_{\theta\theta}\left(\gamma_{\theta\theta}\phi_{\theta\theta}+\nu\gamma_{rr}\phi_{rr}\right), \label{cons-rel-ESK-2} \\
&&M_{rr}=B\gamma_{rr}\left(\gamma_{rr}\phi_{rr} +\nu\gamma_{\theta\theta}\phi_{\theta\theta} \right), \label{cons-rel-ESK-3} \\
&&M_{\theta\theta}=B\gamma_{\theta\theta}\left( \gamma_{\theta\theta} \phi_{\theta\theta} +\nu\gamma_{rr}\phi_{rr} \right). \label{cons-rel-ESK-4}
\end{eqnarray}
\end{subequations}

The comparison between the constitutive relations in
Eqs.~(\ref{cons-rel-Bio}) and Eqs.~(\ref{cons-rel-ESK}) underlines
once again the difference between the two models. While the former
relations are linear, the latter are nonlinear; while in the former
$\sigma_{\alpha\alpha}$ depend only on $\gamma_{\alpha\alpha}$ and
$M_{\alpha\alpha}$ depend only on $\phi_{\alpha\alpha}$, in the
latter there are mixed terms.

A natural question then is how the actual stresses given by these
relations correspond to the ones obtained by variation of the energy
with respect to the strain as it is defined in each model. In the
present model they are identical; compare Eqs.~(\ref{cons-rel-Bio}) to
Eqs.~(\ref{sigma-2D}) and (\ref{M-2D}). This is because the relation
between $\gamma_{\alpha\alpha}$ and the strain
$\epsilon_{\alpha\alpha}$ used in this model is linear; hence,
$\delta\epsilon_{\alpha\alpha}=\delta\gamma_{\alpha\alpha}$. The
stress and moments tensors, $s^{\alpha\alpha}$ and $m^{\alpha\alpha}$,  which were defined in Ref.~\cite{Efi1} differ from the actual ones, Eqs.~(\ref{cons-rel-ESK}). The stress, $s^{\alpha\alpha}$, is based on
variation of the energy with respect to the strain $\tilde\epsilon_{\alpha\alpha}$ and the bending moment, $m^{\alpha\alpha}$, is based on variation of the energy with respect to the second fundamental form $b_{\alpha\alpha}$.
  The two sets of stresses and bending moments, $\{\sigma_{\alpha\alpha},M_{\alpha\alpha}\}$  from Eqs.~(\ref{cons-rel-ESK}) and $\{s^{\alpha\alpha},m^{\alpha\alpha}\}$ are inter-related
according to
\begin{subequations}
\label{ESK-tensors}
\begin{eqnarray}
\mbox{ESK:}\ \ \ \ 
  \sigma_{rr} &=& \gamma_{rr}s^{rr}+\phi_{rr}m^{rr},\label{ESK-tensors-1} \\
  \sigma_{\theta\theta} &=& \Phi^2(\gamma_{\theta\theta}s^{\theta\theta}+\phi_{\theta\theta}m^{\theta\theta}),\label{ESK-tensors-2}
 \\
  M_{rr} &=& \gamma_{rr}m^{rr}, \label{ESK-tensors-3}\\
  M_{\theta\theta} &=& \Phi^2\gamma_{\theta\theta}m^{\theta\theta}.\label{ESK-tensors-4}
\end{eqnarray}
\end{subequations}

In summary, the equations of equilibrium
(\ref{force-torque-balance-2D-2}) and (\ref{stress-torque-com-2}) are
model-independent and, in particular, common to the two models
compared here. They become different only once the different
constitutive relations, either (\ref{cons-rel-Bio}) or
(\ref{cons-rel-ESK}), are substituted in them. Upon this substitution,
one obtains the equations of equilibrium, predicted by the respective
model from minimization of its respective energy over spatial
configurations ${\bf f}$. We now demonstrate it in two examples.

\subsection{Flat deformations}

In the case of flat deformations, $\zeta=0$, we have from both models
, $M_{\alpha\alpha}=0$. Substituting this
result in the torque balance equation
(\ref{force-torque-balance-2D-2}), we get also that the normal
stresses vanishing, $\sigma_{\alpha 3}=0$. In addition, for the case of planar
axisymmetric deformations Eq.~(\ref{force-torque-balance-2D-1}) is
automatically satisfied in the tangential and normal directions. Thus,
the only non-vanishing equation is the balance of forces in the radial
direction, which reads,
\begin{equation}\label{planar-force-bala-app}
\partial_r(\Phi\sigma_{rr})-\sigma_{\theta\theta}=0.
\end{equation} 
This recovers Eq.~(\ref{in-plane-eq-sigma1}) of the main text. Since
we have not yet used a constitutive relation, this equation holds also
in the ESK model.

Substituting in Eq.~(\ref{planar-force-bala-app}) the constitutive
relations of the present model, Eqs.~(\ref{cons-rel-Bio-1}) and
(\ref{cons-rel-Bio-2}), we recover the linear equilibrium equation of
the main text, Eq.~(\ref{eom-flat-ur}).  Repeating the same using the
ESK constitutive relations (\ref{cons-rel-ESK-1}) and
(\ref{cons-rel-ESK-2}), we obtain
\begin{eqnarray}\label{flat-eom-ESK-1}
\partial_r \left(\Phi\gamma_{rr}s^{rr}\right)-\Phi^2\gamma_{\theta\theta}s^{\theta\theta}=0,
\end{eqnarray}
where the ESK stresses of Eq.~(\ref{ESK-tensors}) have been
used. Finally, introducing the Christoffel symbols,
$\Gamma_{rr}^r=\partial_{r}\gamma_{rr}/\gamma_{rr}$ and
$\Gamma_{\theta\theta}^r=-\Phi\gamma_{\theta\theta}/\gamma_{rr}$, we
recover Eq.~(7) of Ref.~\cite{Efi2},
\begin{equation}\label{ESK-eom-7}
\frac{1}{\Phi}\partial_r(\Phi s^{rr})+\Gamma_{rr}^r s^{rr}+\Gamma_{\theta\theta}^r s^{\theta\theta}=0.
\end{equation}
This equation exhibits the covariant form of the ESK theory.  At the
same time it has the disadvantage of being nonlinear in the
displacement, $u_r$, compared to the present model's linear
Eq.~(\ref{eom-flat-ur}).

\subsection{Normal force balance in an isometric immersion}
\label{spherical-cap-ESK}

As a second example we return to the issue addressed in
Sec.~\ref{self-consis}, i.e., the balance of normal forces in the
spherical-cap isometry of a sheet with elliptic reference metric. Once
again, we apply the different sets of constitutive relations of the
two models to the model-independent equations of equilibrium, and
compare the results. In the ESK case this procedure recovers, here
based on force balance, the first of Eqs.~(3.10) in
Ref.~\cite{Efi1}. In the present model it leads to a different
equation of equilibrium. The two equations disagree concerning the
balance of normal forces in an isometric spherical cap, as presented
in Sec.~\ref{self-consis}. The spherical cap satisfies the present
equation and does not satisfy the ESK one. As will be shown below,
this disagreement arises from the additional coupling terms between
stretching and bending appearing in Eqs.~(\ref{cons-rel-ESK-1}) and
(\ref{cons-rel-ESK-2}).

To derive the equation of normal force balance we first project
Eq.~(\ref{force-torque-balance-2D-1}) onto the normal direction, and
Eq.~(\ref{force-torque-balance-2D-2}) onto the tangential direction,
\begin{subequations}\label{normal-torque-isom}
\begin{eqnarray}
0&=&\frac{1}{\Phi}\partial_r(\Phi\sigma_{r3})
+\sigma_{rr}\phi_{rr}+\sigma_{\theta\theta}\phi_{\theta\theta},
\label{normal-torque-isom-1} \\
\gamma_{rr}\Phi\sigma_{r3}&=&-\partial_r(\Phi M_{rr})+M_{\theta\theta}\cos\phi^r.\label{normal-torque-isom-2}
\end{eqnarray}
\end{subequations}
(The geometrical meaning of $\cos\phi^r$ is explained in
Fig.~\ref{epsilon-rr-2theta} of the main text.) Eliminating $\sigma_{r3}$
gives,
\begin{equation}\label{normal-force-balance-fi}
0=\frac{1}{\Phi}\partial_r\left[\frac{1}{\gamma_{rr}}\left(\partial_r(\Phi M_{rr})-M_{\theta\theta}\cos\phi^r\right)\right]-\sigma_{rr}\phi_{rr}
-\sigma_{\theta\theta}\phi_{\theta\theta}.
\end{equation}
Equation~(\ref{normal-force-balance-fi}) expresses normal force
balance regardless of model.

Now, we substitute in Eq.~(\ref{normal-force-balance-fi}) the constitutive
relations of the present model, Eqs.~(\ref{cons-rel-Bio}). For
isometric immersion $\gamma_{rr}=\gamma_{\theta\theta}=1$, we have
from Eqs.~(\ref{cons-rel-Bio-1}) and (\ref{cons-rel-Bio-2}) that 
$\sigma_{\alpha\alpha}=0$. As a result,
Eq.~(\ref{normal-force-balance-fi}) can be integrated, thus
recovering, for free boundary conditions,
Eq.~(\ref{moment-balance-eom}) of the main text. As discussed in
Sec.~\ref{self-consis}, this equation of normal force balance is
satisfied by the spherical cap isometry, Eq.~(\ref{f-shape-elliptic}).

Now we substitute in Eq.~(\ref{normal-force-balance-fi}) the ESK
constitutive relations Eqs.~(\ref{cons-rel-ESK}). This gives
\begin{equation}\label{normal-force-balance-Efi}
0=\frac{1}{\Phi}\partial_r\left[\Phi\left(\frac{1}{\Phi}\partial_r(\Phi m^{rr})+\Gamma_{rr}^r m^{rr}+\Gamma_{\theta\theta}^r m^{\theta\theta}\right)\right]
-s^{rr}b_{rr} -s^{\theta\theta}b_{\theta\theta} - m^{rr}c_{rr}-m^{\theta\theta}c_{\theta\theta},
\end{equation}
where $c_{\alpha\alpha}=\bar{g}_{\alpha\alpha}\phi_{\alpha\alpha}^2$,
and the Christoffel symbols have been used again,
$\Gamma_{rr}^r=\partial_r\gamma_{rr}/\gamma_{rr}$,
$\Gamma_{\theta\theta}^r=-(\Phi\gamma_{\theta\theta}/\gamma_{rr})\cos\phi^r$. 

For the isometry, $\gamma_{\alpha\alpha}=1$, we have from
Eqs.~(\ref{ESK-tensors}) that $M_{rr}=m^{rr}$ and $M_{\theta\theta}=\Phi^2
m^{\theta\theta}$. As a result, the first terms in
Eqs.~(\ref{normal-force-balance-Efi}) and
(\ref{normal-force-balance-fi}) become equal, and the terms
$s^{\alpha\alpha}b_{\alpha\alpha}$ and
$\sigma_{\alpha\alpha}\phi_{\alpha\alpha}$ in the two equations
vanish. However, the last terms in
Eq.~(\ref{normal-force-balance-Efi}),
$m^{\alpha\alpha}c_{\alpha\alpha}$, do not have a counterpart in the
general equation of normal force balance
(\ref{normal-force-balance-fi}). They originate in the bending
contributions appearing in the ESK stresses of
Eqs.~(\ref{ESK-tensors}) or (\ref{cons-rel-ESK}), compared to those of
Eqs.~(\ref{cons-rel-Bio}). They do not vanish for an isometry, leaving
$\sigma_{rr}=\phi_{rr}M_{rr}$ and
$\sigma_{\theta\theta}=\phi_{\theta\theta}M_{\theta\theta}$. Upon
substitution of the spherical cap, Eq.~(\ref{f-shape-elliptic}), in
Eq.~(\ref{normal-force-balance-Efi}), the terms
$m^{\alpha\alpha}c_{\alpha\alpha}$ remain finite, and normal force
balance is not satisfied.

\bibliography{bibitem}{}

\end{document}